\documentclass[aps,preprint,showpacs]{revtex4}
\usepackage{epsfig}
\usepackage{wasysym}
\usepackage{amssymb}

\begin{document}

\newcommand{\bvec}[1]{\mbox{\boldmath ${#1}$}}
\title{Kaon photoproduction in a multipole approach}
\author{T.\ Mart and A. Sulaksono}
\address{Departemen Fisika, FMIPA, Universitas Indonesia, Depok 16424, 
  Indonesia}
\date{\today}
\begin{abstract}
  The recently published experimental data on $K^+\Lambda$ photoproduction 
  by the SAPHIR, CLAS, and LEPS
  collaborations are analyzed by means of a multipole approach. For this
  purpose the background amplitudes are constructed from appropriate 
  Feynman diagrams in a gauge-invariant
  and crossing-symmetric fashion. The results of our calculation emphasize the
  lack of mutual consistency between the SAPHIR and CLAS data previously found
  by several independent research groups, whereas the LEPS data are found 
  to be more consistent with those of 
  CLAS. The use of SAPHIR and CLAS data, individually or simultaneously,
  leads to quite different resonance parameters which, therefore, could
  lead to different conclusions on ``missing resonances''. 
  Fitting to the SAPHIR and LEPS data simultaneously indicates that 
  the $S_{11}(1650)$, $P_{13}(1720)$, $D_{13}(1700)$, 
  $D_{13}(2080)$, $F_{15}(1680)$, and 
  $F_{15}(2000)$ resonances are required, while fitting to the combination
  of CLAS and LEPS data leads alternatively to the $P_{13}(1900)$, $D_{13}(2080)$, 
  $D_{15}(1675)$, $F_{15}(1680)$, and $F_{17}(1990)$ resonances.
  Although yielding 
  different results in most cases, both SAPHIR and CLAS data indicate 
  that the second peak in the cross sections 
  at $W\sim 1900$ MeV originates from the $D_{13}(2080)$ resonance 
  with a mass between 1911 -- 1936 MeV. Furthermore, in contrast to the
  results of currently available models and the Table of Particle Properties,
  both data sets do not exhibit the need for a $P_{11}(1710)$ resonance. The few
  data points available for target asymmetry can not be described by the
  models proposed in the present work. 
\end{abstract}

\maketitle

\section{Introduction}
\renewcommand{\baselinestretch}{1}
Modern theories of the strong interaction would certainly be incomplete if we ignored the
necessity to understand hadronic interactions in the medium energy region. However,
due to the nonperturbative nature of QCD at these energies, hadronic physics
continues to be a challenging field of investigation.
This is also supported by the fact that methods like chiral perturbation
theory are not amenable to this energy region. Lattice QCD, which is 
expected to alleviate this problem, has only recently begun to contribute
to this field.

One of the most intensively studied topics in the realm of hadronic
physics is the associated strangeness photoproduction. High-intensity 
continuous electron beams produced by modern accelerator technologies, 
along with unprecedented precise detectors, are among the important
aspects that have brought renewed attention to this 40 years old field of 
research. On the other hand, the argument that some of 
the resonances predicted by constituent quark
models are strongly coupled to strangeness channels, and therefore
intangible to $\pi N \to \pi N$ reactions that are used by Particle
Data Group (PDG) to extract the properties of nucleon resonances, has raised
the issue of ``missing'' resonances. As a consequence, 
photoproduction of strange particles becomes a unique tool
that can shed important information on the structure of resonances 
and, thus, complement the  $\pi N \to \pi N$ channels. Among the possible
reactions, the $\gamma p\to K^+\Lambda$ is the most intensively studied
channel since it does not involve isospin-3/2 intermediate states
which makes theoretical formalism much simpler. It is also this  
channel for which most of the good quality experimental data are available.
Furthermore, in this process the self-analyzing power of the weak
decay $\Lambda\to p\pi^-$ can be utilized to determine the polarization
of the recoiled $\Lambda$. Therefore, the beauty of working with $K^+\Lambda$
photoproduction is that precise $\Lambda$ polarizations
will accompany accurate cross section measurements. 

In the last decades a large number of attempts have been devoted to
model the above reaction process. Most of these have been performed in the
framework of tree-level isobar models  
\cite{Adelseck:1986fb,Adelseck:1990ch,Williams:1991tw,Han:1999ck,kaon-maid}, 
coupled channel calculations \cite{Feuster:1998cj,Julia-Diaz:2006is,Chiang:2001pw}, 
or quark models \cite{Li:1995sia,Lu:1995bk}. Extending the validity of isobar 
models to higher energy regions has also been recently pursued 
\cite{Mart:2003yb,Mart:2004au,Corthals:2005ce}.

In contrast to pion and eta photoproduction, the kaon photoproduction process 
is not dominated by a single resonant state. Therefore, the main difference 
among the models is chiefly in the use of nucleon,
hyperon, and kaon resonances. The widely used {\small KAON-MAID} model 
\cite{kaon-maid}, for instance, uses nucleon resonances
$S_{11}(1650)$, $P_{11}(1710)$, $P_{13}(1720)$, and $D_{13}(1895)$,
where the latter is known as the missing resonance in this model. 
On the other hand, the Adelseck-Saghai model \cite{Adelseck:1990ch} 
has solely one nucleon resonance $S_{11}(1650)$ and one hyperon resonance 
$S_{01}(1670)$. The more complicated Saclay-Lyon model \cite{David:1995pi} 
utilizes the $P_{11}(1440)$, $P_{13}(1720)$, and $D_{15}(1675)$ nucleon resonances,
along with the $S_{01}(1405)$, $S_{01}(1670)$, $P_{01}(1810)$, and $P_{11}(1660)$
hyperon resonances. Although those models vary with the number of resonances, 
they mostly use low spin states, because higher spin
propagators and vertices are quite complicated in such a 
framework and, moreover, are not free of some fundamental ambiguities.
Only in the Saclay-Lyon \cite{David:1995pi} and Renard-Renard models 
\cite{Renard:1971us} is a spin-5/2 nucleon resonance 
utilized. Other models argue that the use of resonance excitation  
up to spin 3/2, or even up to spin 1/2, is sufficient. 

Clearly, there is a lack of systematic procedure to determine how many
resonances should be built into the process. There has been no attempt
to include the $F_{15}$, $F_{17}$, $G_{17}$, and $G_{19}$ states, although
some of them could have sizable branching fractions to the $K\Lambda$ 
channel (see Table \ref{tab:resonance_pdg} in the next section or
Review of Particle Properties \cite{Eidelman:2004wy}).

The main motivation of the present work is to explore the
possibility of using higher spin states in kaon photoproduction.
Ideally, this should be performed on the basis of a 
coupled-channels formalism. However, the level of complexity
in such a framework increases quickly with the addition of 
resonance states. In view of this, we constrain the present work
to a single-channel analysis, but we use as much as possible nucleon
resonances listed by PDG. This argument is also supported by the
fact that the recently available SAPHIR \cite{Glander:2003jw} 
and CLAS \cite{Bradford:2005pt} data have a problem of
mutual consistency \cite{Bydzovsky:2006wy,Julia-Diaz:2006is}. 
Thus, another purpose of this work is to investigate the physics 
consequence of using each data set. 
The present work is basically an extension
of our previous analysis \cite{Mart:2004ug} which was performed using a 
slightly different method and only the SAPHIR data \cite{Glander:2003jw}.
To this end, we will use the same formalism developed for pion photoproduction 
\cite{hanstein99,Tiator:2003uu} which has the advantage that it provides
a direct comparison of the extracted helicity photon coupling with
the PDG values and paves the way for extending the present work
to include the effect of other channels. 

This paper is organized as follows. In Section \ref{sec:formalism} we present
the formalism of our work. Section \ref{sec:data_and_fitting} briefly 
discusses the experimental data used in the fitting process as well as
the chosen fitting strategy. In Section \ref{sec:results} we discuss the 
comparison of the results of our calculation with the current available data. 
In this section we also discuss the possible
origin of the second peak in the cross sections at $W\sim 1900$ MeV.
In Section \ref{sec:conclusion} we summarize our findings.

\section{Formalism}\label{sec:formalism}
\subsection{The Background Amplitudes}
\label{subsect:background}
The background amplitudes are obtained from a series of tree-level Feynman 
diagrams \cite{background}. 
They consist of the standard $s$-, $u$-, and $t$-channel Born terms 
along with the $K^*(892)$ and $K_1(1270)$ $t$-channel vector mesons. 
Altogether they are often called extended Born terms.
Apart from the $K_1(1270)$ exchange, these background terms are
similar to the ones used by Thom \cite{thom1966}. The importance of the 
$K_1(1270)$ intermediate state has been pointed out for the first time 
by Ref.~\cite{Adelseck:yv} and since then it has been extensively used in 
almost all isobar models. To account for hadronic structures of 
interacting baryons and mesons we include the appropriate hadronic form factors 
in the hadronic vertices by utilizing the method developed by Haberzettl in order 
to maintain gauge invariance of the amplitudes. We have also tested 
the gauge method proposed by Ohta \cite{ohta89}, but since the produced $\chi^2$
is substantially larger, we will not discuss
Ohta method here. Furthermore, to comply 
with the \textit{crossing symmetry} requirement we use a special
form factor in the {\it gauge terms}
\begin{eqnarray}
  \label{eq:fhat2}
  \widehat{F}(s,t,u) &=& F_1(s)+F_1(u)+F_3(t)-F_1(s)F_1(u)\nonumber\\ 
                     && -F_1(s)F_3(t) - F_1(u)F_3(t) + F_1(s)F_1(u)F_3(t) ~,
\end{eqnarray}
proposed by Davidson and Workman \cite{Davidson:2001rk}, with
Mandelstam variables $s$, $t$, and $u$, and
\begin{eqnarray}
  \label{eq:hadr_ff}
  F_i(x)&=& \frac{\Lambda^4}{\Lambda^4+(x-m_i^2)^2} ~,
\end{eqnarray}
where $\Lambda$ and $m_i$ are the form factor cut-off and the
intermediate state mass, respectively \cite{Haberzettl:1998eq}.
Thus, comparing to the previous pioneering work \cite{thom1966}, the
major improvement in the background sector is the use of hadronic 
form factors in a gauge-invariant fashion and the crossing-symmetric
properties of the Born terms.

\subsection{The Resonance Amplitudes}
\label{subsect:multipole}
The resonant electric and magnetic multipoles for a state with the mass $M_R$, width 
$\Gamma$, and angular momentum $\ell$ are assumed to have the Breit-Wigner form
\cite{hanstein99,Tiator:2003uu}
\begin{eqnarray}
  \label{eq:em_multipole}
  A_{\ell\pm}^R(W) &=& {\bar A}_{\ell\pm}^R \, c_{KY}\, \frac{f_{\gamma R}(W)\, 
    \Gamma_{\rm tot}(W) M_R\, f_{K R}(W)}{M_R^2-W^2-iM_R\Gamma_{\rm tot}(W)}~ e^{i\phi} ~,
  \label{eq:m_multipole}
\end{eqnarray}
where $W$ represents the total c.m. energy, the isospin factor 
$c_{KY}$ is $-1$ \cite{Chiang:2001as}, $f_{KR}$ is the usual
Breit-Wigner factor describing the decay of a resonance $R$ with a total width
$\Gamma_{\rm tot}(W)$ and physical mass $M_R$. The $f_{\gamma R}$ indicates
the $\gamma NR$ vertex and $\phi$ represents the phase angle. 
The Breit-Wigner factor $f_{KR}$ is given by 
\begin{eqnarray}
  \label{eq:f_KR}
  f_{KR}(W) &=& \left[\frac{1}{(2j+1)\pi}\frac{k_W}{|\bvec{q}|}\frac{m_N}{W}
  \frac{\Gamma_{KY}}{\Gamma_{\rm tot}^2}\right]^{1/2}~~,~~~~
  k_W ~=~ \frac{W^2-m_N^2}{2W}~,
\end{eqnarray}
with $m_N$ the nucleon mass. The energy dependent partial width $\Gamma_{KY}$ 
is defined through
\begin{eqnarray}
  \label{eq:Gamma_KY}
  \Gamma_{KY} &=& \beta_K\Gamma_R \left(\frac{|\bvec{q}|}{q_R}\right)^{2\ell+1}
  \,\left(\frac{X^2+q_R^2}{X^2+\bvec{q}^2}\right)^\ell\frac{W_R}{W}~,
\end{eqnarray}
where the damping parameter $X$ is assumed to be 500 MeV for all resonances,
$\beta_K$ is the single kaon branching ratio, 
$\Gamma_R$ and $q_R$ are the total width and kaon c.m. momentum at $W=M_R$. 
The $\gamma NR$ vertex is parameterized through
\begin{eqnarray}
  \label{eq:f_gammaR}
  f_{\gamma R} &=& \left(\frac{k_W}{k_R}\right)^{2\ell '+1}\,\left(\frac{X^2+k_R^2}{
    X^2+k_W^2}\right)^{\ell '} ~,
\end{eqnarray}
where $k_R$ is equal to $k_W$ calculated at $W=M_R$. For $M_{\ell\pm}$ and $E_{\ell +}$: 
$\ell' =\ell$, whereas for $E_{\ell -}$: $\ell '= \ell -2$ if $\ell \ge 2$ \cite{Aznauryan:2002gd}.
The values
of $\ell$ and $\ell '$ for all resonances considered in this study are given in 
Table~\ref{tab:multipoles}.

The total width appearing in Eqs.~(\ref{eq:m_multipole}) and (\ref{eq:f_KR}) is the 
sum of $\Gamma_K$ and the ``inelastic'' width $\Gamma_{\rm in}$. In this work
we assume the dominance of the pion decay channel and we 
parameterize the width by using
\begin{eqnarray}
  \label{eq:Gamma_tot}
  \Gamma_{\rm tot} ~=~ \Gamma_{KY}+\Gamma_{\rm in}~,~~~~
  \Gamma_{\rm in} ~=~ (1-\beta_K) \Gamma_R \left(\frac{q_\pi}{q_0}\right)^{2\ell+4}
    \left(\frac{X^2+q_0^2}{X^2+q_\pi^2}\right)^{\ell+2}~,
\end{eqnarray}
with $q_\pi$ the momentum of the $\pi$ in the decay of $R\to\pi+N$ in c.m.
system and $q_0=q_\pi$ calculated at $W=M_R$.

The electric and magnetic multipole photon couplings ${\bar A}_{\ell\pm}^R$ 
in Eq.\,(\ref{eq:em_multipole}) can be related to the helicity photon couplings
$A_{1/2}$ and $A_{3/2}$. For resonances with total spin $j=\ell+1/2$ we get
\cite{Tiator:2003uu}
\begin{eqnarray}
  \label{eq:helicity1}
  A_{1/2}^{\ell+} &=& -\frac{1}{2}\,\left[(\ell+2){\bar E}_{\ell+}+ \ell{\bar M}_{\ell+}\right]~,\\
  A_{3/2}^{\ell+} &=& \frac{1}{2}\,\sqrt{\ell(\ell+2)}\left({\bar E}_{\ell+}-{\bar M}_{\ell+}\right)~,
\end{eqnarray}
and for $j=(\ell+1)-1/2$
\begin{eqnarray}
  A_{1/2}^{(\ell+1)-} &=& \frac{1}{2}\,\left[(\ell+2){\bar M}_{(\ell+1)-}- \ell{\bar E}_{(\ell+1)-}\right]~,\\
  A_{3/2}^{(\ell+1)-} &=& -\frac{1}{2}\,\sqrt{\ell(\ell+2)}\left[{\bar E}_{(\ell+1)-}+{\bar M}_{(\ell+1)-}\right]~.
  \label{eq:helicity2}
\end{eqnarray}
Equations (\ref{eq:helicity1})--(\ref{eq:helicity2}) can be inverted to give the electric
and magnetic multipole photon couplings in terms of the helicity photon couplings as
\begin{eqnarray}
  \label{eq:El+}
  {\bar E}_{\ell+} &=& \frac{1}{\ell+1}\left[ -A_{1/2}^{\ell+}+\sqrt{\frac{\ell}{
        \ell+2}}\, A_{3/2}^{\ell+}\right]~,\\
  {\bar M}_{\ell+} &=& -\frac{1}{\ell+1}\left[ A_{1/2}^{\ell+}+\sqrt{\frac{\ell+2}{
        \ell}}\, A_{3/2}^{\ell+}\right]~,
  \label{eq:Ml+}
\end{eqnarray}
for $j=\ell+1/2$, and
\begin{eqnarray}
  \label{eq:E(l+1)-}
  {\bar E}_{(\ell+1)-} &=& -\frac{1}{\ell+1}\left[ A_{1/2}^{(\ell+1)-}+\sqrt{\frac{\ell+2}{
        \ell}}\, A_{3/2}^{(\ell+1)-}\right]~,\\
  {\bar M}_{(\ell+1)-} &=& \frac{1}{\ell+1}\left[ A_{1/2}^{(\ell+1)-}-\sqrt{\frac{\ell}{
        \ell+2}}\, A_{3/2}^{(\ell+1)-}\right]~,
  \label{eq:M(l+1)-}
\end{eqnarray}
for $j=(\ell+1)-1/2$. All relevant multipole photon couplings used in this work
are given in Table~\ref{tab:multipoles}.

\begin{table}[t]
  \centering
  \caption{The electric and magnetic multipole photon couplings in terms of the helicity 
    photon couplings for resonances up to $\ell=4$.}
  \label{tab:multipoles}
  \begin{ruledtabular}
  \begin{tabular}[c]{ccccc}
    Resonance & $\ell$ & $\ell '$ & Multipoles & Expression \\
    \hline
    $S_{11}$ &0&0& ${\bar E}_{0+}$ & $-A_{1/2}^{0+}$ \\
    $P_{11}$ &1&1& ${\bar M}_{1-}$ & $A_{1/2}^{1-}$ \\
    $P_{13}$ &1&1& ${\bar E}_{1+}$ & $\frac{1}{2}\left[-A_{1/2}^{1+}+\sqrt{\frac{1}{3}}A_{3/2}^{1+}\right]$\\
             &1&1& ${\bar M}_{1+}$ & $-\frac{1}{2}\left[A_{1/2}^{1+}+\sqrt{3}A_{3/2}^{1+}\right]$\\
    $D_{13}$ &2&0& ${\bar E}_{2-}$ & $-\frac{1}{2}\left[A_{1/2}^{2-}+\sqrt{3}A_{3/2}^{2-}\right]$\\
             &2&2& ${\bar M}_{2-}$ & $\frac{1}{2}\left[A_{1/2}^{2-}-\sqrt{\frac{1}{3}}A_{3/2}^{2-}\right]$\\
    $D_{15}$ &2&2& ${\bar E}_{2+}$ & $\frac{1}{3}\left[-A_{1/2}^{2+}+\sqrt{\frac{1}{2}}A_{3/2}^{2+}\right]$\\
             &2&2& ${\bar M}_{2+}$ & $-\frac{1}{3}\left[A_{1/2}^{2+}+\sqrt{2}A_{3/2}^{2+}\right]$\\
    $F_{15}$ &3&1& ${\bar E}_{3-}$ & $-\frac{1}{3}\left[A_{1/2}^{3-}+\sqrt{2}A_{3/2}^{3-}\right]$\\
             &3&3& ${\bar M}_{3-}$ & $\frac{1}{3}\left[A_{1/2}^{3-}-\sqrt{\frac{1}{2}}A_{3/2}^{3-}\right]$\\
    $F_{17}$ &3&3& ${\bar E}_{3+}$ & $\frac{1}{4}\left[-A_{1/2}^{3+}+\sqrt{\frac{3}{5}}A_{3/2}^{3+}\right]$\\
             &3&3& ${\bar M}_{3+}$ & $-\frac{1}{4}\left[A_{1/2}^{3+}+\sqrt{\frac{5}{3}}A_{3/2}^{3+}\right]$ \\
    $G_{17}$ &4&2& ${\bar E}_{4-}$ & $-\frac{1}{4}\left[A_{1/2}^{4-}+\sqrt{\frac{5}{3}}A_{3/2}^{4-}\right]$\\
             &4&4& ${\bar M}_{4-}$ & $\frac{1}{4}\left[A_{1/2}^{4-}-\sqrt{\frac{3}{5}}A_{3/2}^{4-}\right]$ \\
    $G_{19}$ &4&4& ${\bar E}_{4+}$ & $\frac{1}{5}\left[-A_{1/2}^{4+}+\sqrt{\frac{2}{3}}A_{3/2}^{4+}\right]$\\
             &4&4& ${\bar M}_{4+}$ & $-\frac{1}{5}\left[A_{1/2}^{4+}+\sqrt{\frac{3}{2}}A_{3/2}^{4+}\right]$ \\[1ex]
  \end{tabular}
  \end{ruledtabular}
\end{table}

\begin{table}[tb]
  \centering
  \caption{Resonances up to $\ell=4$ with the corresponding properties from the 
  Review of Particle Properties \cite{Eidelman:2004wy}.}
  \label{tab:resonance_pdg}
  \begin{ruledtabular}
  \begin{tabular}[c]{cccccccc}
    Resonance & $M_R$ & $\Gamma_R$ & $\beta_K$ & $A_{1/2} (p)$ & $A_{3/2}(p)$ & Overall & Status\\
     & (MeV) & (MeV) && ($10^{-3}$ GeV$^{-1/2}$) & ($10^{-3}$ GeV$^{-1/2}$) & status & seen in $K\Lambda$\\
    \hline
    $S_{11}$ & 1650 & 150 &$0.027\pm 0.004$& $+53\pm 16$ & - & **** & ***\\
    & 2090 & 400 &-& - & - & * & -\\
    $P_{11}$ & 1710 & 100 &$0.050\pm 0.020$& $ +9\pm 22$ & - & *** & **\\
    & 2100 & 200 &-& - & - & * & -\\
    $P_{13}$ & 1720& 150 &-& $+18\pm 30$ & $-19\pm 20$ & **** & **\\
    & 1900& 498 &$0.001\pm 0.001$& - & - & ** & -\\
    $D_{13}$ & 1700& 100 &-& $-18\pm 13$ & $-2\pm 24$ & *** & **\\
    & 2080& 450 &$0.002\pm 0.002$& $-20\pm 8$ & $17\pm 11$ & ** & *\\
    $D_{15}$ & 1675 & 150 &-& $+19\pm 8$ & $15\pm 9$ & **** & *\\
             & 2200& 130 &-& - & - & ** & *\\
    $F_{15}$ & 1680 & 130 &-& $-15\pm 6$ & $133\pm 12$ & **** & -\\
             & 2000& 490 &-& - & - & ** & *\\
    $F_{17}$ & 1990 & 535 &-& $+30\pm 29$ & $86\pm 60$ & ** & *\\
    $G_{17}$ & 2190 & 450 &-& $-55$ & +81 & **** & *\\
    $G_{19}$ & 2250  & 400 &-& - & - & **** & -\\
  \end{tabular}
  \end{ruledtabular}
\end{table}

All observables can be calculated from the CGLN amplitudes \cite{Knochlein:1995qz}
\begin{eqnarray}
  \label{eq:cgln}
  {F} &=& i \bvec{\sigma}\cdot\bvec{\epsilon}\, F_1 + 
  \bvec{\sigma}\cdot\hat{\bvec{q}}\,\bvec{\sigma}\cdot (\hat
  {\bvec{k}}\times\bvec{\epsilon})\, F_2 + i\bvec{\sigma}\cdot
  \hat{\bvec{k}}\,\hat{\bvec{q}}\cdot \bvec{\epsilon}\, F_3 +
  i\bvec{\sigma}\cdot\hat{\bvec{q}}\,\hat{\bvec{q}}\cdot \bvec{\epsilon}\, F_4 ~,~~~
\end{eqnarray}
where the amplitudes $F_i$ are related to the electric and magnetic multipoles
given in Eq.\,(\ref{eq:em_multipole}) for up to $\ell=4$ by
\begin{eqnarray}
  \label{eq:f1}
  F_1 &=& E_{0+}-{\textstyle \frac{3}{2}}(E_{2+}+2M_{2+})+
          E_{2-}+3M_{2-}\nonumber\\
      & & +{\textstyle \frac{15}{8}}(E_{4+}+4M_{4+})
      - {\textstyle \frac{3}{2}}(E_{4-}+5M_{4-}) \nonumber\\
      & & +\,3\left\{ E_{1+}+M_{1+}-{\textstyle \frac{5}{2}}(E_{3+}+3M_{3+})
            +E_{3-}+4M_{3-}\right\}\, \cos\theta \nonumber\\
      & & + \,{\textstyle \frac{15}{2}}\left\{ E_{2+}+2M_{2+} -
          {\textstyle \frac{7}{2}}(E_{4+}+4M_{4+}) + 
          E_{4-}+5M_{4-}\right\}\,\cos^2\theta\nonumber\\
      & & +\, {\textstyle \frac{35}{2}}(E_{3+}+3M_{3+})\,\cos^3\theta
          + {\textstyle \frac{315}{8}}(E_{4+}+4M_{4+})\,\cos^4\theta ~,\\
  \label{eq:f2}
  F_2 &=& 2M_{1+}+M_{1-}-{\textstyle \frac{3}{2}}(4M_{3+}+3M_{3-})
          \nonumber\\
      & & + 3\left\{3M_{2+}+2M_{2-} - {\textstyle \frac{5}{2}}
          (5M_{4+}+4M_{4-})\right\}\,\cos\theta \nonumber\\
      & & + {\textstyle \frac{15}{2}}(4M_{3+}+3M_{3-})\cos^2\theta
          + {\textstyle \frac{35}{2}}(5M_{4+}+4M_{4-})\cos^3\theta ~,\\
  \label{eq:f3}
  F_3 &=& 3\left\{\, E_{1+}-M_{1+}-{\textstyle \frac{5}{2}}(E_{3+}-M_{3+})+
          E_{3-}+M_{3-}\right\} \nonumber\\
      & & +\, 15\left\{\, E_{2+}-M_{2+}+E_{4-}+M_{4-}-
          {\textstyle \frac{7}{2}}(E_{4+}-M_{4+})\right\}\,\cos\theta\nonumber\\
      & & +\, {\textstyle \frac{105}{2}}(E_{3+}-M_{3+})\,\cos^2\theta + 
          {\textstyle \frac{315}{2}}(E_{4+}-M_{4+})\cos^3\theta ~,\\
  \label{eq:f4}
  F_4 &=& 3\left\{\, M_{2+}-E_{2+}-M_{2-}-E_{2-}- {\textstyle \frac{5}{2}}
          (M_{4+}-E_{4+}-M_{4-}-E_{4-})\right\}\nonumber\\
      & & + 15 (M_{3+}-E_{3+}-M_{3-}-E_{3-})\,\cos\theta
          + {\textstyle \frac{105}{2}}(M_{4+}-E_{4+}\nonumber\\
      & & -M_{4-}-E_{4-})\,\cos^2\theta ~.
\end{eqnarray}
These amplitudes are combined with the CGLN amplitudes obtained from the background
terms discussed in Subsection \ref{subsect:background}. 

\section{Experimental Data and Fitting Strategy}\label{sec:data_and_fitting}
\subsection{Experimental Data}
All experimental data used in the present analysis are summarized in 
Table~\ref{tab:exp_data}. The 2004 SAPHIR differential cross section
data are given in 36 angular distributions with about 20 MeV energy bins,
spanning from reaction threshold up to $W=2.4$ GeV. The obtained differential
cross sections are represented in 20 panels showing energy distributions. The
recoil polarization is also given in 5 angular bins.

The latest version of the CLAS data \cite{Bradford:2005pt} covers almost the
same energy range, from threshold up to about 2.5 GeV, but are given in
76 bins of angular distributions, with a step size in $E_\gamma$ of 25 MeV,
or about 9 - 14 MeV in $W$, covering 18 angular bins in $W$ excitations. The
recoil polarization data are taken from the previous data analysis 
\cite{McNabb2004}, which are presented in 29 angular distributions.
In addition, both SAPHIR and CLAS present figures of total cross sections,
but do not give tables of numerical values. In our analysis we did not 
use these data in the fits, but we show these for the sake of comparison.

The LEPS collaboration reported the SPRING8 data in terms of differential
cross sections and photon asymmetries at forward kaon angles
\cite{Sumihama:2005er}. These data are used in our analysis.

In addition, we will also compare the three data points from an old measurement 
of the target asymmetry \cite{Althoff:1978qw} with our results. 
These data have been recently used by several authors 
\cite{Julia-Diaz:2006is,Maxwell:2004ga,Mart:1999ed,David:1995pi}.

\begin{table}[tb]
  \centering
  \caption{Experimental data sets used in the present analysis.
  Experimental data set used in the individual fits are indicated by $\surd$.
  Otherwise, data are only used for comparison.}
  \label{tab:exp_data}
  \begin{ruledtabular}
  \begin{tabular}[c]{llcrcccc}
    Name & Observable & Symbol& $N$~ & Fit 1 & Fit 2 & Fit 3& Ref. \\
    \hline
    SAPHIR 2004 & Differential cross section &$d\sigma/d\Omega$&720&$\surd$&-&$\surd$& \cite{Glander:2003jw}\\
    & Recoil polarization & $P$ &30&$\surd$&-&$\surd$& \cite{Glander:2003jw}\\
    & Total cross section & $\sigma_{\rm tot}$ & 36 & - & - & - & \cite{Glander:2003jw}\\
    CLAS 2006 &Differential cross section &$d\sigma/d\Omega$ &1377&-&$\surd$&$\surd$&\cite{Bradford:2005pt}\\
    & Recoil polarization & $P$ &233&-&$\surd$&$\surd$& \cite{Bradford:2005pt}\\
    & Total cross section & $\sigma_{\rm tot}$ & 78 & - & - & - & \cite{Bradford:2005pt}\\
    LEPS 2006 &Differential cross section &$d\sigma/d\Omega$ & 54&$\surd$&$\surd$&$\surd$&\cite{Sumihama:2005er}\\
     &Photon asymmetry &$\Sigma$ &30 &$\surd$&$\surd$&$\surd$&\cite{Sumihama:2005er}\\
     OLD & Target asymmetry & $T$ & 3 & - & - & - & \cite{Althoff:1978qw}\\
     & Total cross section & $\sigma_{\rm tot}$ & 24 & - & - & - & \cite{Tran:1998qw}\\
     \hline
     Total data & & & &834&1694&2444& \\
  \end{tabular}
  \end{ruledtabular}
\end{table}

Since several recent studies have reported the problem of mutual consistency between
SAPHIR and CLAS data \cite{Bydzovsky:2006wy,Julia-Diaz:2006is}, 
in our database we define three different data sets,
as shown in Table~\ref{tab:exp_data}. In the first fit (Fit 1) we use only SAPHIR
and LEPS data, while in the second one (Fit 2) we use a combination of 
CLAS and LEPS data. In the last one (Fit 3) we use all data (SAPHIR, CLAS, and LEPS)
in the fit.

Note that we do not use the old data in any fit, since Ref.~\cite{Bydzovsky:2006wy}
has pointed out that some of them (old SAPHIR data) are consistent only with the new
SAPHIR data, while some others (much older measurements \cite{older-measurment}) 
are consistent only with the CLAS data.

\subsection{Fitting Strategy}
Recent analyses of kaon photoproduction have mostly focused on the quest
of missing resonances. With the new CLAS data appearing this year
\cite{Bradford:2005pt}, this becomes an arduous task, since 
Ref.~\cite{Bydzovsky:2006wy} found a lack of mutual consistency between
the recent CLAS and SAPHIR data. As will be shown in the next section,
the use of the two data sets, individually or simultaneously, leads to 
quite different values of the extracted resonance parameters and, therefore,
could yield different conclusions on the missing resonances studied by this
reaction. In view of this, in the present work we do not focus our 
attention on searching for
missing resonances. Instead, we will use all nucleon resonances listed
by PDG up to spin 9/2 and fit their parameters to new data. Along with 
their known parameters those resonances are listed in 
Table~\ref{tab:resonance_pdg}. Note that we do not use resonances with
masses below the reaction threshold (1610 MeV) since their contributions 
would only contribute to the background terms and, therefore, would 
be difficult to see in the present formalism. Furthermore, we 
do not include the two resonances with spin higher than $9/2$ (i.e.,
$I_{1\, 11}$ and $K_{1\, 13}$) for practical reasons and because
too little information is available for both states.

The number of free parameters is relatively large, i.e., 7 from the
background amplitude and 86 from the resonance part. To reduce this
we fix both $g_{K\Lambda N}$ and $g_{K\Sigma N}$ coupling constants
to the SU(3) predictions, i.e., $g_{K\Lambda N}/\sqrt{4\pi}
=-3.80$ and $g_{K\Sigma N}/\sqrt{4\pi}=1.20$, and fix masses as well as widths of the
four-star resonances to their PDG values. To avoid unrealistically
large values obtained from fitting to experimental data, the total width
$\Gamma_R$ is limited to 500 MeV and the kaon branching ratio $\beta_K$
is limited to 0.3. The $\chi^2$ minimization fit is performed 
by using the CERN-MINUIT code.

\section{Results and Discussion}\label{sec:results}
\subsection{Numerical Results}

Contributions to the $\chi^2$ from individual data for all fits are given in 
Table~\ref{tab:chi}. By comparing the contributions for Fit 1 and Fit 2 
it is clear that the LEPS data are more compatible with the CLAS rather 
than with the SAPHIR measurement. It is also apparent that, 
in spite of the large number of data, the CLAS differential cross sections 
(74\%) are internally more consistent than the SAPHIR ones (84\%). 
These results corroborate the finding of Ref.~\cite{Bydzovsky:2006wy}.
Of course, the situation significantly changes 
when all data are simultaneously used in the fit, i.e.,
since the CLAS error bars are in general smaller than those of SAPHIR, 
the CLAS differential cross sections contribute more to the $\chi^2$
than the SAPHIR ones.

\begin{table}[tb]
  \centering
  \caption{Contribution to $\chi^2$ (in \%) from individual data sets for the three different fits.}
  \label{tab:chi}
  \begin{ruledtabular}
  \begin{tabular}[c]{llrccc}
    Name & Observable & $N$~ & Fit 1 & Fit 2 & Fit 3 \\
    \hline
    SAPHIR 2004 & Differential cross section &720& 84 &- & 39 \\
    & Recoil polarization & 30& 3 &- & 1 \\
    CLAS 2006 &Differential cross section &1377& - & 74 & 45 \\
    & Recoil polarization & 233& - & 17 & 9 \\
    LEPS 2006 &Differential cross section  & 54& 10 & 7 & 5 \\
     &Photon asymmetry  &30 & 3 & 2 & 1 \\
  \end{tabular}
  \end{ruledtabular}
\end{table}

\begin{figure*}[t]
\centering
 \mbox{\epsfig{file=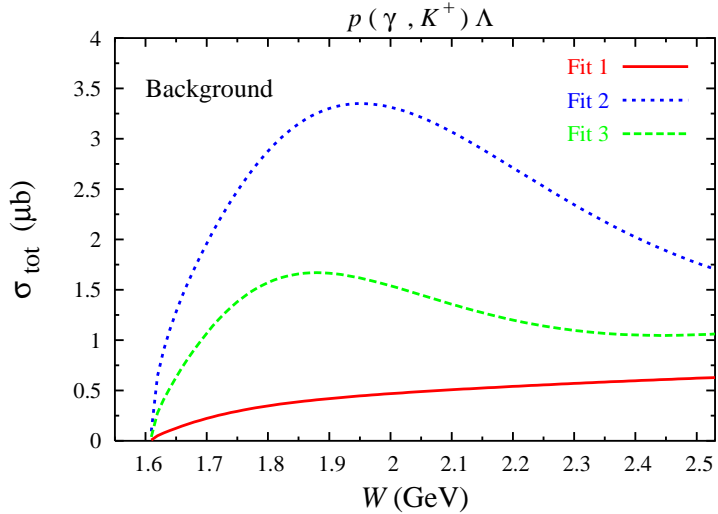,width=10cm}}
\caption{(Color online) Contributions from the background amplitudes
  to the total cross sections for all fits.}\label{fig:born}
\end{figure*}

\begin{figure*}[!]
\centering
 \mbox{\epsfig{file=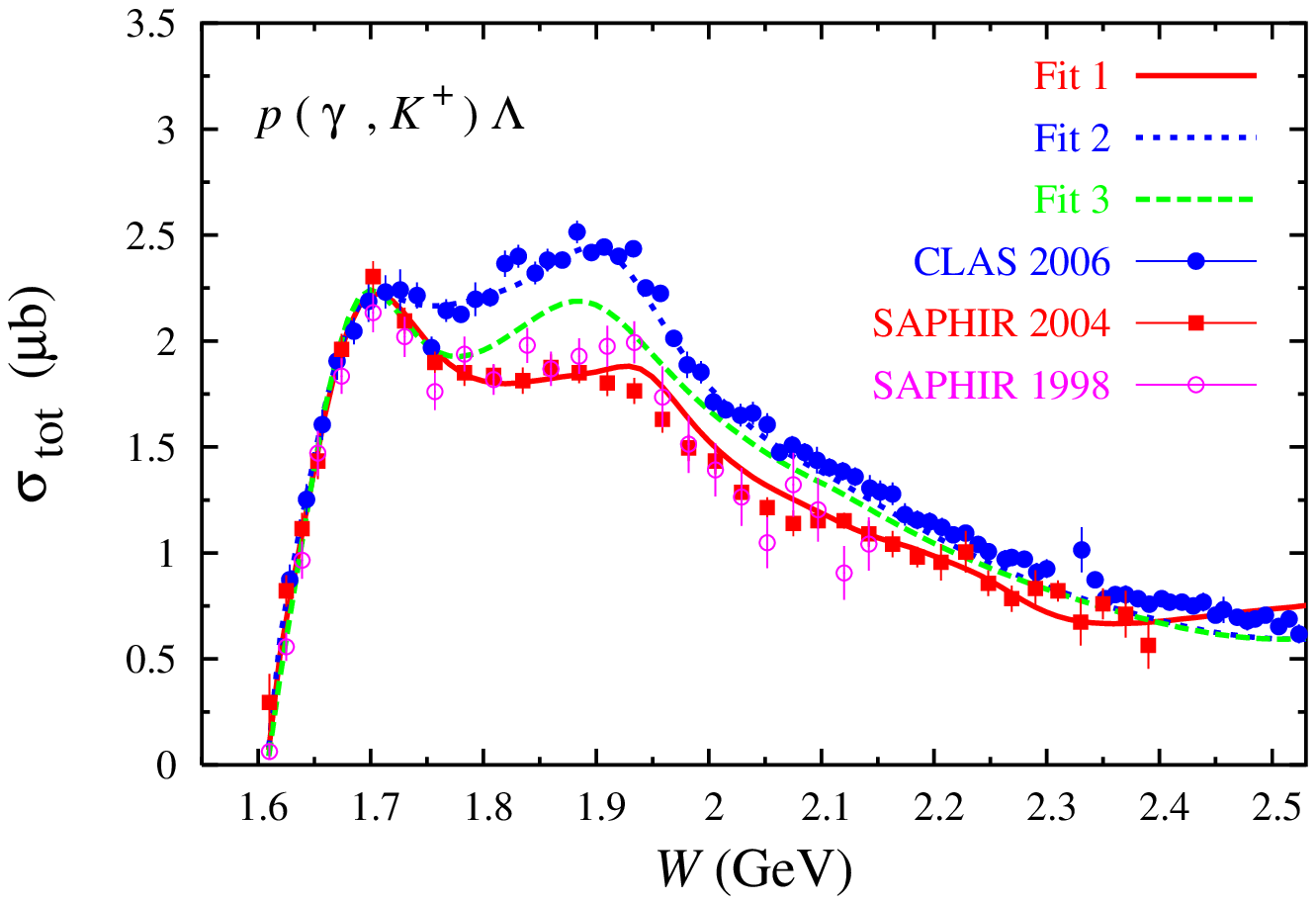,width=10cm}}
\caption{(Color online) Comparison between the calculated total cross sections
  with experimental data. All data shown in this figure
  were not used in the fits.}\label{fig:total}
\end{figure*}

The extracted background parameters are shown in Table~\ref{tab:results_cc}.
In this case it is interesting to note the different values in the 
vector mesons coupling constants extracted from different sets of data. 
Obviously, fitting to the CLAS and LEPS data results in smaller coupling 
constants. However, the corresponding hadronic form factor cut-off is
significantly larger than that obtained in Fit 1. Including all data sets
in the database leads to a compromise result, i.e., the extracted 
parameters basically lie between those obtained from Fit 1 and Fit 2.
To investigate the qualitative effect of these parameters 
on the observable, in Fig.\,\ref{fig:born} we compare contributions of
the background terms to the total cross sections of all fits. Obviously,
the different values of coupling constants lead to different size
backgrounds. For Fit 1, the large coupling constants combined with 
the soft form factor cut-off yields a small background which systematically increases 
as a function of $W$. On the contrary, the relatively hard hadronic form factor
cut-off in Fit 2 is unable to suppress the large contribution of the 
standard Born terms and, therefore, yields a large background. Although tending 
to be convergent at higher energies, such a large background seems  
to be unrealistic if we compare it to experimental data shown in Fig.\,\ref{fig:total}.
This indicates that the extracted hadronic form factor cut-off (i.e.,
$\Lambda=1.13$ GeV) is presumably too large for kaon photoproduction. A similar 
result has been found in the isobar model of Ref.~\cite{Mart:1999ed},
where a value of $\Lambda=0.80$ GeV is demanded. 
As expected, a compromise background will be 
obtained if we use all data sets. 

Since in general the error bars of CLAS data are smaller than those of SAPHIR data, 
the smaller value of $\chi^2/N_{\rm dof}$ exhibited by 
Fit 2 in Table~\ref{tab:results_cc} again indicates that the CLAS data
show a better internal consistency than the SAPHIR data, 
a point which has been discussed previously.

\begin{table}[b]
  \centering
  \caption{The extracted coupling constants, hadronic form factor cut-offs ($\Lambda$),
  number of data ($N$), and $\chi^2$ per number of degrees of freedom from the three 
  different fits.}
  \label{tab:results_cc}
  \begin{ruledtabular}
  \begin{tabular}[c]{ccccccrrc}
    & ${\displaystyle \frac{G_V(K^*)}{4\pi}}$ 
    & ${\displaystyle \frac{G_T(K^*)}{4\pi}}$ & ${\displaystyle \frac{G_V(K_1)}{4\pi}}$ & ${\displaystyle \frac{G_T(K_1)}{4\pi}}$ & $\Lambda$ (GeV) & $N$ &$\chi^2$& $\chi^2/N_{\rm dof}$\\
    [1em]
    \hline
    Fit 1 &$~~ 1.19\pm 0.19$ &$-3.57\pm 0.48$ &$ 0.18\pm 0.32$ &$-4.95\pm 1.23$ &$ 0.50\pm 0.02$ &   834&772& 1.02 \\
    Fit 2 &$~~ 0.06\pm 0.00$ &$-0.18\pm 0.01$ &$ 0.16\pm 0.01$ &$-1.13\pm 0.02$ &$ 1.13\pm 0.01$ &  1694&1581& 0.98 \\ 
    Fit 3 &$-0.02\pm 0.01$ &$-0.64\pm 0.04$ &$ 0.71\pm 0.04$ &$-1.94\pm 0.07$ &$ 0.91\pm 0.02$ &  2444&3095& 1.31 \\
  \end{tabular}
  \end{ruledtabular}
\end{table}

The extracted resonance parameters for all three different fits are shown in 
Table\,\ref{tab:results_resonance}. The agreements with the PDG values 
are mostly good or fair and for some resonances there are significant 
discrepancies. Since in this framework (single channel analysis)
we are unable to consider the effect of other channels
which could substantially influence the widths, photon couplings, and branching
fractions of those resonances, we will only qualitatively discuss the variation 
of certain resonance parameters obtained for the three different fits. 
A more detailed study, in which these parameters are constrained within
the PDG values and all information from other related channels (i.e.,
$\pi N, \pi\pi N, \eta N$) is taken into account in the inelastic width
of Eq.~(\ref{eq:Gamma_tot}), will be reported in a future publication
\cite{mart-tiator}.

\begin{table}[!]
  \centering
  \caption{The extracted resonance parameters from the three different fits.
  Values written in italic were fixed during the fitting process.}
  \label{tab:results_resonance}
  \begin{ruledtabular}
  \begin{tabular}[c]{rccccccc}
    Resonance & $M_R$ & $\Gamma_R$ & $A_{1/2}$ & $A_{3/2}$& $\beta_K$ & $\phi$ & $\Delta\chi^2$\\
    (status) & (MeV) & (MeV) & ($10^{-3}$ GeV$^{-1/2}$) & ($10^{-3}$ GeV$^{-1/2}$) & & (deg.)& (\%) \\
    \hline
$S_{11}$(1650)  Fit 1 & {\sl 1650} & {\sl 150} & $  27  \pm  1  $ &   -  & $0.300\pm 0.033$ & $ 228\pm   8$ &  12.1 \\
                Fit 2 & {\sl 1650} & {\sl 150} & $   3  \pm  1  $ &   -  & $0.300\pm 0.283$ & $ 119\pm  22$ &   0.1 \\
                Fit 3 & {\sl 1650} & {\sl 150} & $  23  \pm  1  $ &   -  & $0.300\pm 0.001$ & $ 192\pm   4$ &   2.2 \\
        (****)  PDG   &  1650 &  150 &$53\pm 16$&  -  & 0.027 &   -  & - \\
\hline 
$S_{11}$(2090)  Fit 1 & $ 2261\pm  15$ & $ 241\pm  33$ & $  39  \pm  2 $ &   -  & $0.096\pm 0.010$ & $  55\pm   7$ &   5.0 \\
                Fit 2 & $ 2411\pm  12$ & $ 377\pm  25$ & $  63  \pm  4 $ &   -  & $0.044\pm 0.005$ & $  37\pm   5$ &   5.6 \\
                Fit 3 & $ 2492\pm  10$ & $ 500\pm   9$ & $  46  \pm  3 $ &   -  & $0.173\pm 0.022$ & $  81\pm   3$ &   4.5 \\
           (*)  PDG   &  2090 &  400 &   -  &   -  &    -  &   -  & - \\
\hline 
$P_{11}$(1710)  Fit 1 & $ 1709\pm  13$ & $ 150\pm  56$ & $  22  \pm  6 $ &   -  & $0.019\pm 0.010$ & $ 189\pm  16$ &   2.0 \\
                Fit 2 & $ 1720\pm   3$ & $ 150\pm   5$ & $  98  \pm  4 $ &   -  & $0.010\pm 0.052$ & $ 191\pm   2$ &   2.5 \\
                Fit 3 & $ 1720\pm   2$ & $ 150\pm  39$ & $  30  \pm  3 $ &   -  & $0.029\pm 0.008$ & $ 183\pm   4$ &   2.3 \\
         (***)  PDG   &  1710 &  100 &$9\pm 22$&   -  & 0.050 &   -  & - \\
\hline 
$P_{11}$(2100)  Fit 1 & $ 2129\pm  19$ & $  90\pm  20$ & $  -3  \pm  1  $ &   -  & $0.289\pm 0.215$ & $ 244\pm  40$ &   0.8 \\
                Fit 2 & $ 2102\pm   4$ & $  90\pm   4$ & $   5  \pm  1  $ &   -  & $0.300\pm 0.152$ & $   0\pm   0$ &   3.8 \\
                Fit 3 & $ 2104\pm   4$ & $  90\pm  12$ & $  11  \pm  3  $ &   -  & $0.029\pm 0.014$ & $   0\pm   5$ &   1.5 \\
           (*)  PDG   &  2100 &  200 &   -  &   -  &    -  &   -  & - \\
\hline 
$P_{13}$(1720)  Fit 1 &  {\sl 1720} & {\sl 150} & $  -22  \pm  1 $ & $ -20  \pm  2  $ & $0.300\pm 0.053$ & $  46\pm    3$ &   8.4 \\
                Fit 2 &  {\sl 1720} & {\sl 150} & $  -54  \pm  3 $ & $ -49  \pm  2  $ & $0.097\pm 0.008$ & $ 136\pm   2$ &   4.7 \\
                Fit 3 &  {\sl 1720} & {\sl 150} & $  -31  \pm  3 $ & $ -22  \pm  2  $ & $0.248\pm 0.042$ & $  84\pm    3$ &   7.1 \\
        (****)  PDG   &  1720 &  150 &$18\pm 30$& $-19\pm 20$ &    -  &   -  & - \\
\hline 
$P_{13}$(1900)  Fit 1 & $ 1937\pm    7$ & $ 102\pm  24$  & $   43  \pm  5 $ & $ -24  \pm 13 $ & $0.011\pm 0.001$ & $ 240\pm   12$ &   4.4 \\
                Fit 2 & $ 1800\pm   5$ & $ 500\pm  15$   & $   69  \pm  2 $ & $  88  \pm  1 $ & $0.203\pm 0.007$ & $ 208\pm   1$ &   6.1 \\
                Fit 3 & $ 1818\pm   12$ & $ 363\pm   29$ & $  194  \pm  5 $ & $  52  \pm 10 $ & $0.011\pm 0.001$ & $ 165\pm    6$ &   2.9 \\
          (**)  PDG   &  1900 &  498 &  -   &   -  & 0.001 &   -  & - \\
\hline 
$D_{13}$(1700)  Fit 1 & $ 1680\pm    3$ & $ 170\pm  30$  & $  13  \pm  1  $ & $ 18  \pm  2 $ & $0.300\pm 0.027$ & $ 110\pm    8$ &   9.7 \\
                Fit 2 & $ 1750\pm  45$ & $ 500\pm 385$   & $  56  \pm  4  $ & $ 93  \pm  4 $ & $0.010\pm 0.060$ & $  36\pm   2$ &   3.6 \\
                Fit 3 & $ 1682\pm    3$ & $ 499\pm   20$ & $  42  \pm  2  $ & $ 69  \pm  3 $ & $0.010\pm 0.001$ & $  37\pm    2$ &   1.7 \\
         (***)  PDG   &  1700 &  100 &$-18\pm 13$& $-2\pm 24$ &   -   &   -  & - \\
\hline 
$D_{13}$(2080)  Fit 1 & $ 1936\pm   10$& $ 301\pm  22$ & $ -26  \pm  2$ & $-32  \pm  3$ & $0.300\pm 0.063$ & $  54\pm    7$ &   8.8 \\
                Fit 2 & $ 1915\pm   4$ & $ 165\pm   8$ & $-140  \pm  7$ & $ 32  \pm  4$ & $0.012\pm 0.001$ & $   1\pm   3$ &  8.5  \\
                Fit 3 & $ 1911\pm    4$& $ 193\pm   9$ & $ -85  \pm  2$ & $ 28  \pm  3$ & $0.034\pm 0.002$ & $   8\pm    3$ &   7.1 \\
          (**)  PDG   &  2080 &  450 &$-20\pm 8$& $17\pm 11$ & 0.002 &   -  & - \\
  \end{tabular}
  \end{ruledtabular}
\end{table}

\begin{table}[!]
  \addtocounter{table}{-1}
  \centering
  \caption{The extracted resonance parameters from the three different fits (continued).}
  \begin{ruledtabular}
  \begin{tabular}[c]{rccccccc}
    Resonance & $M_R$ & $\Gamma_R$ & $A_{1/2}$ & $A_{3/2}$& $\beta_K$ & $\phi$ & $\Delta\chi^2$\\
    (status) & (MeV) & (MeV) & ($10^{-3}$ GeV$^{-1/2}$) & ($10^{-3}$ GeV$^{-1/2}$) & & (deg.)& (\%)\\
    \hline
$D_{15}$(1675)  Fit 1 & {\sl 1675} & {\sl 150} & $ 22  \pm  3$ & $ 30  \pm  3 $ & $0.010\pm 0.008$ & $ 243\pm    7$ &   4.2 \\
                Fit 2 & {\sl 1675} & {\sl 150} & $ -2  \pm  0$ & $-15  \pm  1 $ & $0.164\pm 0.029$ & $ 212\pm   3$ &  7.4  \\
                Fit 3 & {\sl 1675} & {\sl 150} & $  0  \pm  1$ & $-13  \pm  1 $ & $0.226\pm 0.015$ & $ 203\pm    3$ &   3.3 \\
        (****)  PDG   &  1675 &  150 &$19\pm 8$& $15\pm 9$ &   -   &   -  & - \\
\hline 
$D_{15}$(2200)  Fit 1 & $ 2247\pm   13$ & $  90\pm  10$ & $-15  \pm  2$ & $  3  \pm  4$ & $0.026\pm 0.006$ & $ 108\pm   29$ &   1.7 \\
                Fit 2 & $ 2299\pm 177$ & $ 358\pm  41$  & $-13  \pm  4$ & $-31  \pm  5$ & $0.038\pm 0.012$ & $  39\pm  11$ &   2.6 \\
                Fit 3 & $ 2125\pm   17$ & $ 500\pm   34$& $-41  \pm  2$ & $-19  \pm  3$ & $0.050\pm 0.007$ & $   6\pm    5$ &   3.4 \\
          (**)  PDG   &  2200 &  130 &  -   &   -  &   -   &   -  & - \\
\hline 
$F_{15}$(1680)  Fit 1 & {\sl 1680} & {\sl 130} & $-11  \pm  4$ & $11  \pm  2 $ & $0.010\pm 0.014$ & $ 190\pm    7$ &   5.8 \\
                Fit 2 & {\sl 1680} & {\sl 130} & $-17  \pm  1$ & $15  \pm  1 $ & $0.010\pm 0.006$ & $   5\pm   2$ &   6.0 \\
                Fit 3 & {\sl 1680} & {\sl 130} & $-11  \pm  1$ & $25  \pm  1 $ & $0.011\pm 0.001$ & $  27\pm    3$ &   3.4 \\
        (****)  PDG   &  1680 &  130 &$-15\pm 6$& $133\pm 12$ &   -   &   -  & - \\
\hline 
$F_{15}$(2000)  Fit 1 & $ 2100\pm    9$ & $ 345\pm  69$ & $116  \pm 25$ & $ 36  \pm 12 $ & $0.010\pm 0.001$ & $ 208\pm    6$ &   6.8 \\
                Fit 2 & $ 1937\pm   4$ & $ 153\pm  10$  & $ 53  \pm  4$ & $-14  \pm  3 $ & $0.031\pm 0.005$ & $   0\pm  17$ &   4.5 \\
                Fit 3 & $ 1967\pm    7$ & $ 213\pm   15$& $ 33  \pm  3$ & $-61  \pm  5 $ & $0.020\pm 0.002$ & $  55\pm    5$ &   4.5 \\
          (**)  PDG   &  2000 &  490 &  -   &   -  &   -   &   -  & - \\
\hline 
$F_{17}$(1990)  Fit 1 & $ 1970\pm   15$ & $ 169\pm  30$ & $-21  \pm  4$ & $-19  \pm  5$ & $0.040\pm 0.010$ & $  61\pm    9$ &   2.7 \\
                Fit 2 & $ 2083\pm  15$ & $ 531\pm  44$  & $-12  \pm  1$ & $-15  \pm  1$ & $0.300\pm 0.285$ & $  72\pm   3$ &   8.0 \\
                Fit 3 & $ 2056\pm   17$ & $ 394\pm   48$& $-17  \pm  1$ & $-10  \pm  1$ & $0.233\pm 0.027$ & $  45\pm    6$ &   4.7 \\
          (**)  PDG   &  1990 &  535 &$30\pm 29$& $86\pm60$ &   -   &   -  & - \\
\hline 
$G_{17}$(2190)  Fit 1 & {\sl 2190} & {\sl 450} & $-6  \pm  2$ & $-14  \pm  3$ & $0.300\pm 0.252$ & $  27\pm  252$ &   2.1 \\
                Fit 2 & {\sl 2190} & {\sl 450} & $-7  \pm  1$ & $ 14  \pm  1$ & $0.300\pm 0.272$ & $  10\pm   3$ &   4.9 \\
                Fit 3 & {\sl 2190} & {\sl 450} & $-7  \pm  1$ & $ 26  \pm  3$ & $0.102\pm 0.022$ & $   6\pm    5$ &   3.8 \\
        (****)  PDG   &  2190 &  450 &$-55$& $81$ &   -   &   -  & - \\
\hline 
$G_{19}$(2250)  Fit 1 & {\sl 2250} & {\sl 400} & $-20  \pm  8$ & $-18  \pm 11$ & $0.020\pm 0.018$ & $ 153\pm   13$ &   1.6 \\
                Fit 2 & {\sl 2250} & {\sl 400} & $  6  \pm  1$ & $  3  \pm  1$ & $0.300\pm 0.227$ & $  75\pm   7$ &   1.9 \\
                Fit 3 & {\sl 2250} & {\sl 400} & $  9  \pm  1$ & $ -1  \pm  1$ & $0.300\pm 0.194$ & $ 100\pm    5$ &   3.7 \\
        (****)  PDG   &  2250 &  400 &  -  &   -  &   -   &   -  & - \\
  \end{tabular}
  \end{ruledtabular}
\end{table}

The first resonance which is of interest is the $P_{11}(1710)$, for which the extracted
masses and widths from all fits are in agreement with the PDG values. In the case
of Fit 1 the extracted helicity photon coupling is also in agreement with the 
PDG $A_{1/2}$ value. 
The second resonance is the $P_{11}(2100)$, for which fitting to 
SAPHIR or CLAS data leads to a mass which is in good agreement with the PDG 
value, whereas, however, the extracted width underpredicts the PDG value by
almost 50\%. In the case of Fit 2 this resonance is required 
to explain a small bump in the cross sections around 2150 MeV (see panels for 
forward angles in Fig.~\ref{fig:clas_dkpl_e}).
This bump seems to be very tiny in the SAPHIR differential cross section data, but it 
is still visible in the total cross sections (see Fig.~\ref{fig:total}).
The third resonance is the $D_{13}(2080)$, for which all three fits 
clearly underestimate the PDG mass. This will be discussed in 
the next subsection. The fourth resonance is the $F_{15}(2000)$.
In this case SAPHIR data demand this resonance to explain the small peak at 
$W\sim 2150$ MeV, whereas CLAS data require it to describe the second peak
around $W\sim 1900$ MeV. The last resonance is the $F_{17}(1990)$ for which
SAPHIR data lead to resonance parameters which are in a fair agreement 
(up to the sign of the photon couplings) with PDG values, 
whereas CLAS data overestimate the PDG mass. 

\begin{figure}[t]
\centering
 \mbox{\epsfig{file=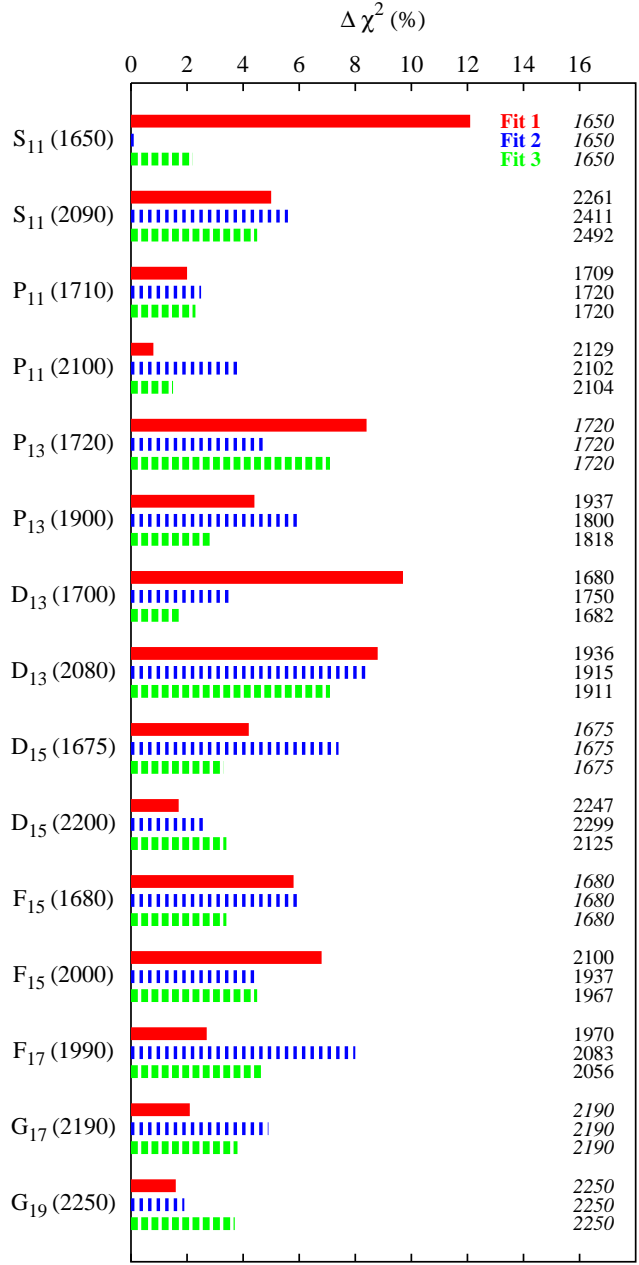,angle=-90,width=9cm}}
\caption{(Color online) The significance of individual resonances in the three
  different fits. Values written in italic were fixed during
  the fit process.}\label{fig:strength}
\end{figure}

To further investigate the importance of the individual resonances we define a parameter
\begin{eqnarray}
  \label{eq:par_res}
  \Delta \chi^2 ~=~ \frac{\chi^2_{\rm All}-\chi^2_{{\rm All}-N^*}}{\chi^2_{\rm All}}
  \times 100\,\% ~,
\end{eqnarray}
where $\chi^2_{\rm All}$ is the $\chi^2$ obtained by using all resonances 
and $\chi^2_{{\rm All}-N^*}$ is the $\chi^2$ obtained by using all 
but a specific resonance. Therefore, $\Delta \chi^2$ 
measures the relative difference between the $\chi^2$ of including and of 
excluding the corresponding resonance. Note that the $\Delta \chi^2$ does not
measure the ``strength'' of the resonance in the process but it merely reveals 
information on how difficult to reproduce experimental data without that 
resonance. A similar ratio has been also defined in Ref.~\cite{Julia-Diaz:2006is}
in order to investigate the role of individual resonances. 
The numerical result is listed in the last column of 
Table\,\ref{tab:results_resonance}.
However, the result would be more clear if displayed in a histogram shown
in Fig.~\ref{fig:strength}. Except for the $S_{11}(2090)$, $P_{11}(1710)$, 
$D_{13}(2080)$, $F_{15}(1680)$, and $G_{19}(2250)$, for which 
the $\Delta\chi^2$ are almost similar, the histogram shows that 
the new CLAS and SAPHIR data can be only explained by different sets of 
nucleon resonances. If we, for instance, trivially define the important 
resonances as those with $\Delta\chi^2 \gtrsim 6\%$, in other words we 
pick about 30\% out of all resonances used in both Fit 1 and Fit 2, 
then the important resonances in Fit 1 are the $S_{11}(1650)$, 
$P_{13}(1720)$, $D_{13}(1700)$, $D_{13}(2080)$, $F_{15}(1680)$, 
and $F_{15}(2000)$, whereas Fit 2 requires the $P_{13}(1900)$, $D_{13}(2080)$, 
$D_{15}(1675)$, $F_{15}(1680)$, and $F_{17}(1990)$. However, fitting all
data simultaneously yields a compromise result and change this conclusion which 
indicates that the corresponding result 
is neither consistent with Fit 1 nor with Fit 2. 

It is interesting to note here
that both Fit 1 and Fit 2 support the requirement of the 
$D_{13}(2080)$ in the $K^+\Lambda$ production process. Surprisingly,
all new data reject the need for the $P_{11}(1710)$, and the new CLAS data 
do not require the contribution from the $P_{13}(1720)$ resonance. 
However, most recent analyses of the $K^+\Lambda$ channel have included these 
intermediate states. Another new phenomenon is the contribution 
from $F_{15}(2000)$ and $F_{17}(1900)$, which are quite important 
according to SAPHIR and CLAS data, respectively.
These resonances have not been used in most analyses, especially in the isobar
model with diagrammatic technique, since propagators for spins 5/2 and 7/2 
are not only quite complicated in this approach, but also their forms 
are not unique. 

Except for the $F_{17}(1990)$ resonance, the importance of individual resonances
discussed above is generally confirmed by the reasonable error bars of the
fitted parameters shown in Table \ref{tab:results_resonance}. In the case of
$F_{17}(1990)$ the error bar of the kaon branching ratio $\beta_K$ is almost the
same as the value of $\beta_K$ itself, whereas, on the other hand, the corresponding
$\Delta\chi^2$ indicates that this resonance is strongly needed to explain
the CLAS data. We have tried to understand this by relaxing the upper limit
of $\beta_K$ and refitting the $F_{17}(1990)$ resonance parameters. It is found
that with the same value of $\chi^2$ the extracted $\beta_K$ is 
$0.387\pm 0.150$, which indicates that this resonance is still important
for Fit 2.

\begin{figure}[!]
\centering
 \mbox{\epsfig{file=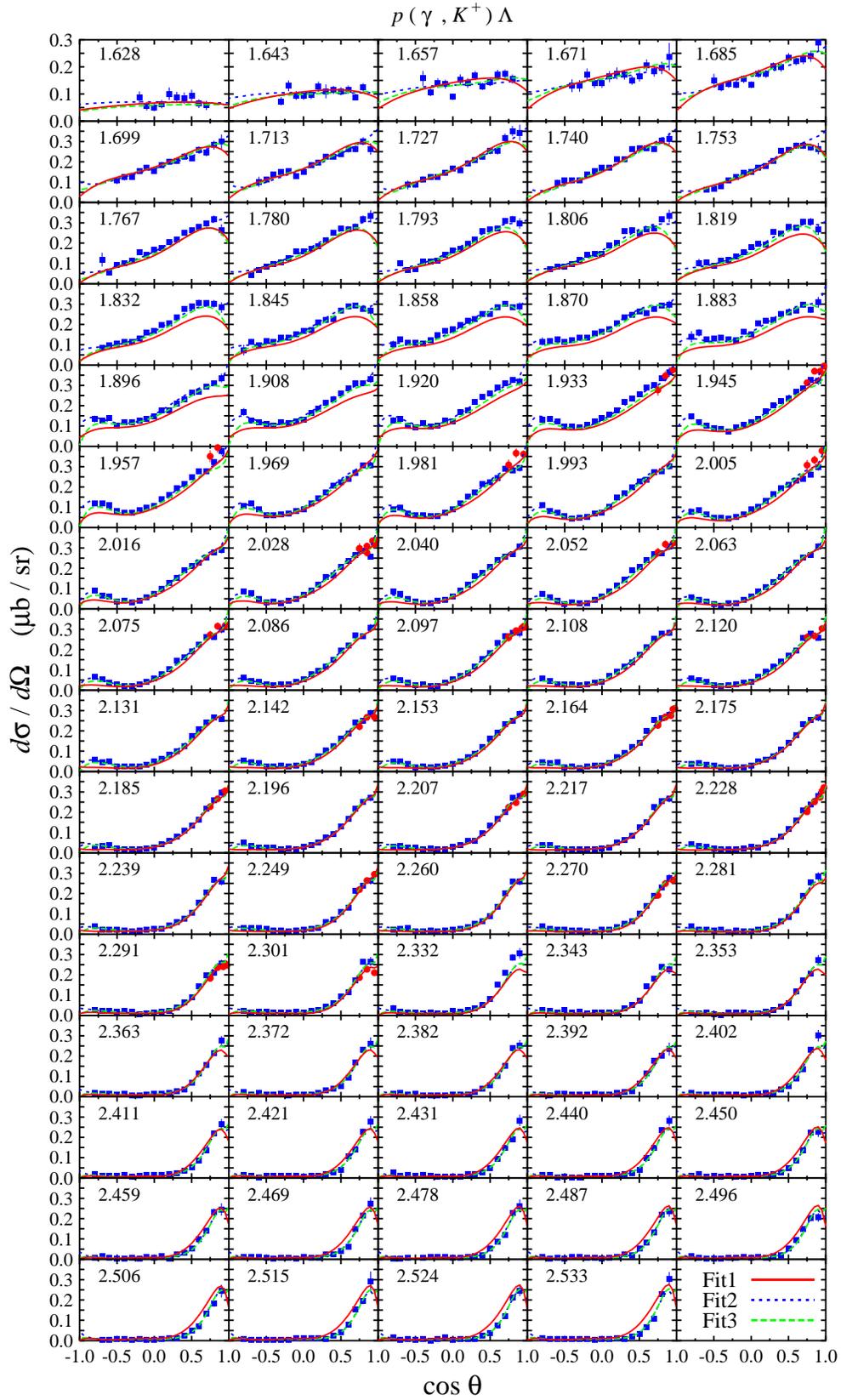,width=13.5cm}}
 \vspace{-2mm}
\caption{(Color online) Comparison between angular distribution of differential cross sections 
  obtained from the three fits with CLAS (solid squares) and LEPS 
  (solid circles) data.}\label{fig:clas_dkpl}
\end{figure}   

\begin{figure}[ht]
\centering
 \mbox{\epsfig{file=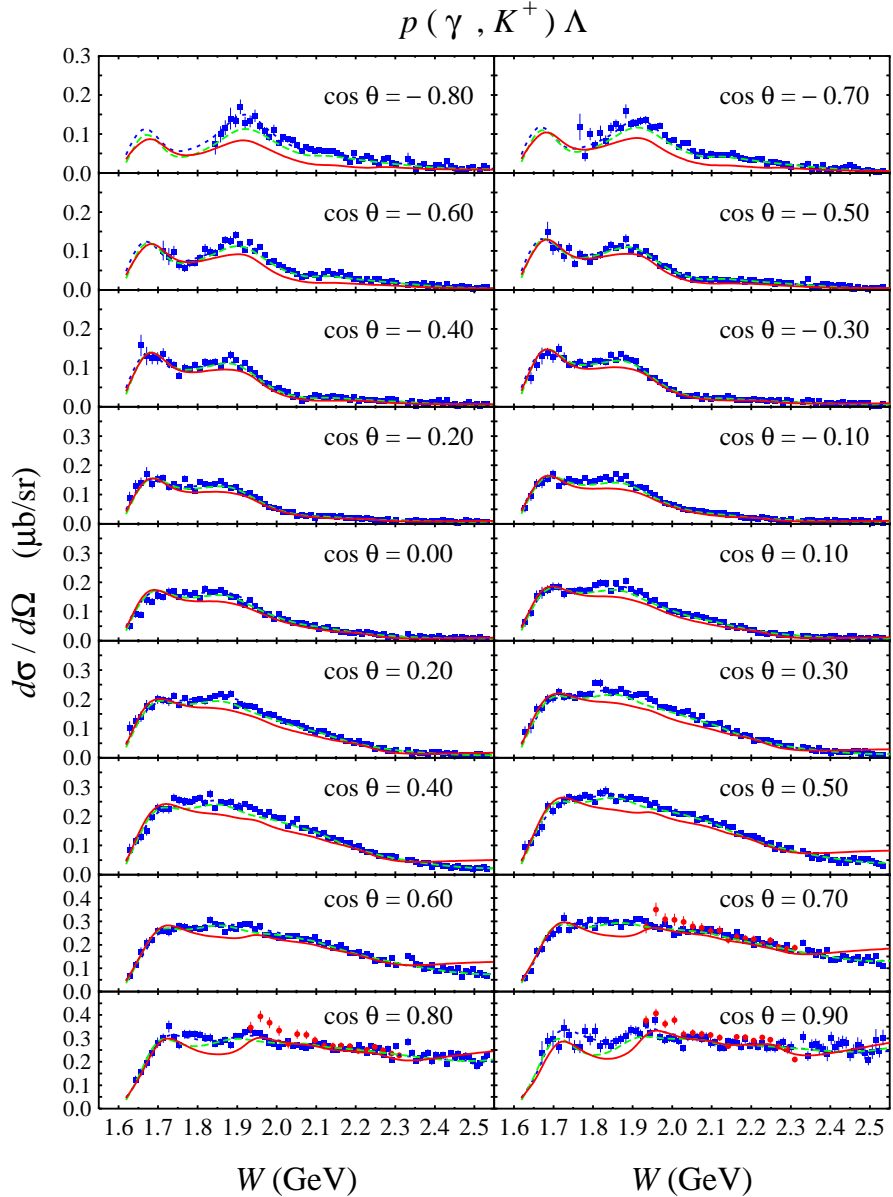,width=12cm}}
\caption{(Color online) Same as in Fig\,\ref{fig:clas_dkpl}, but for the energy distribution.}
\label{fig:clas_dkpl_e}
\end{figure}   

\begin{figure}[!]
\centering
 \mbox{\epsfig{file=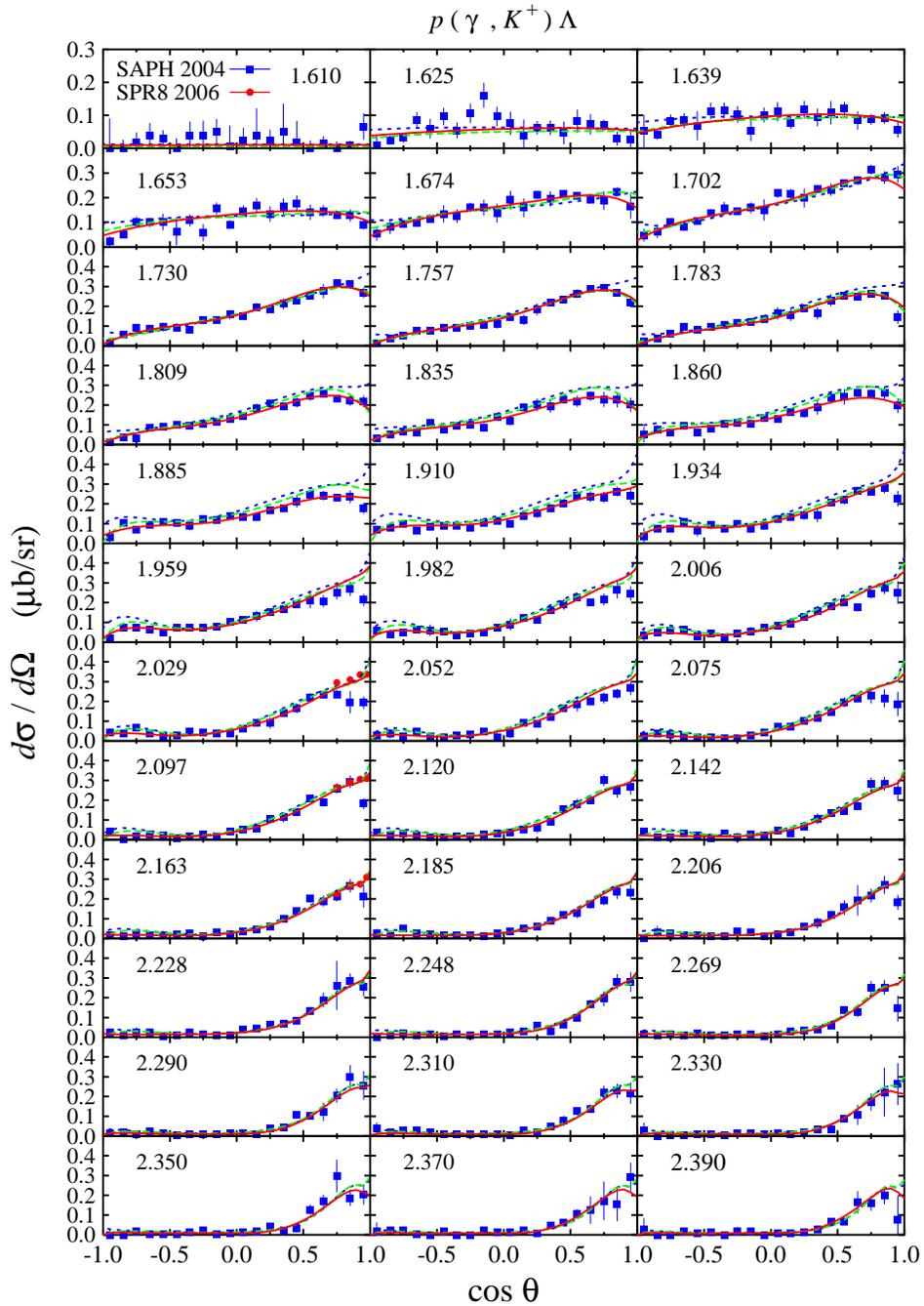,width=13cm}}
\caption{(Color online) Comparison between angular distribution of differential cross sections 
  obtained from the three fits with SAPHIR and LEPS data. Notation for the curves 
  is the same as in Fig.~\ref{fig:clas_dkpl}.}\label{fig:saph_dkpl}
\end{figure}

\begin{figure}[!]
\centering
 \mbox{\epsfig{file=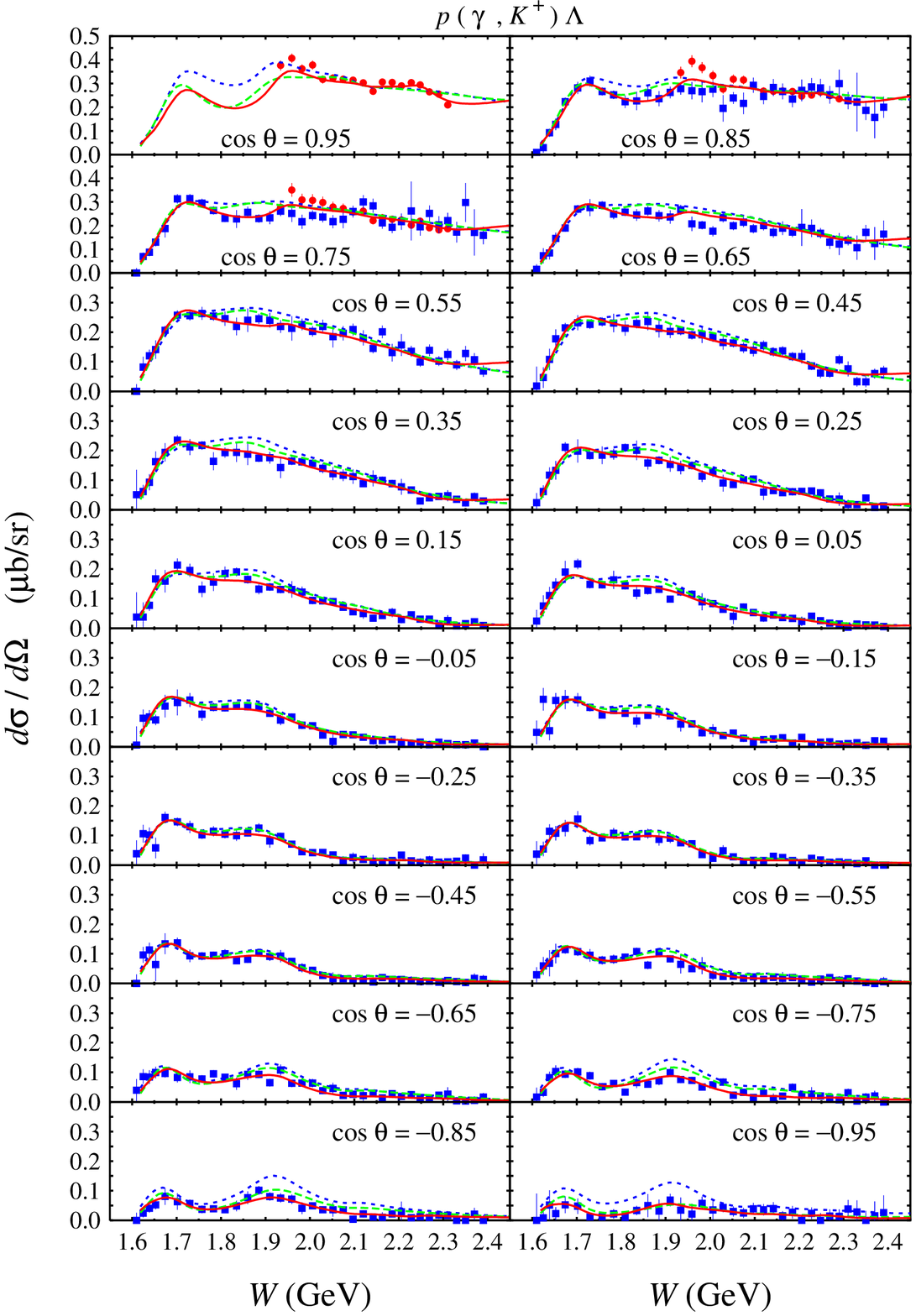,width=12cm}}
\caption{(Color online) Same as in Fig\,\ref{fig:saph_dkpl}, but for the energy distribution.}
\label{fig:saph_dkpl_e}
\end{figure}

Since the CLAS and SAPHIR data are binned in different energy and angular bins,
a simultaneous comparison of the results with both data sets 
in one figure can not be performed. 
In Figs.~\ref{fig:clas_dkpl} and \ref{fig:clas_dkpl_e} we show the comparison
between predictions from all fits with the CLAS data, while the comparisons with
SAPHIR data are shown in Figs.~\ref{fig:saph_dkpl} and \ref{fig:saph_dkpl_e}.
The LEPS data are shown in both cases. It is obvious from those figures that
the LEPS data are more consistent to the CLAS data than with the SAPHIR
measurement. This emphasizes the previous discussion on the numerical result
of Table~\ref{tab:chi}. From the four figures it is also clear that the 
largest discrepancy appears between $W=1.75$ GeV and 1.95 GeV in the forward
direction, whereas in the backward direction the discrepancies show up
in a wider range, i.e., from 1.8 to 2.4 GeV. It is also important to note that
at the very forward and backward angles the two data sets (also,
as a consequence, Fit 1 and Fit 2) exhibit very different trends. The CLAS data
tend to rise at these regions, while the SAPHIR data tend to decrease. 

\begin{figure}[!]
\centering
 \mbox{\epsfig{file=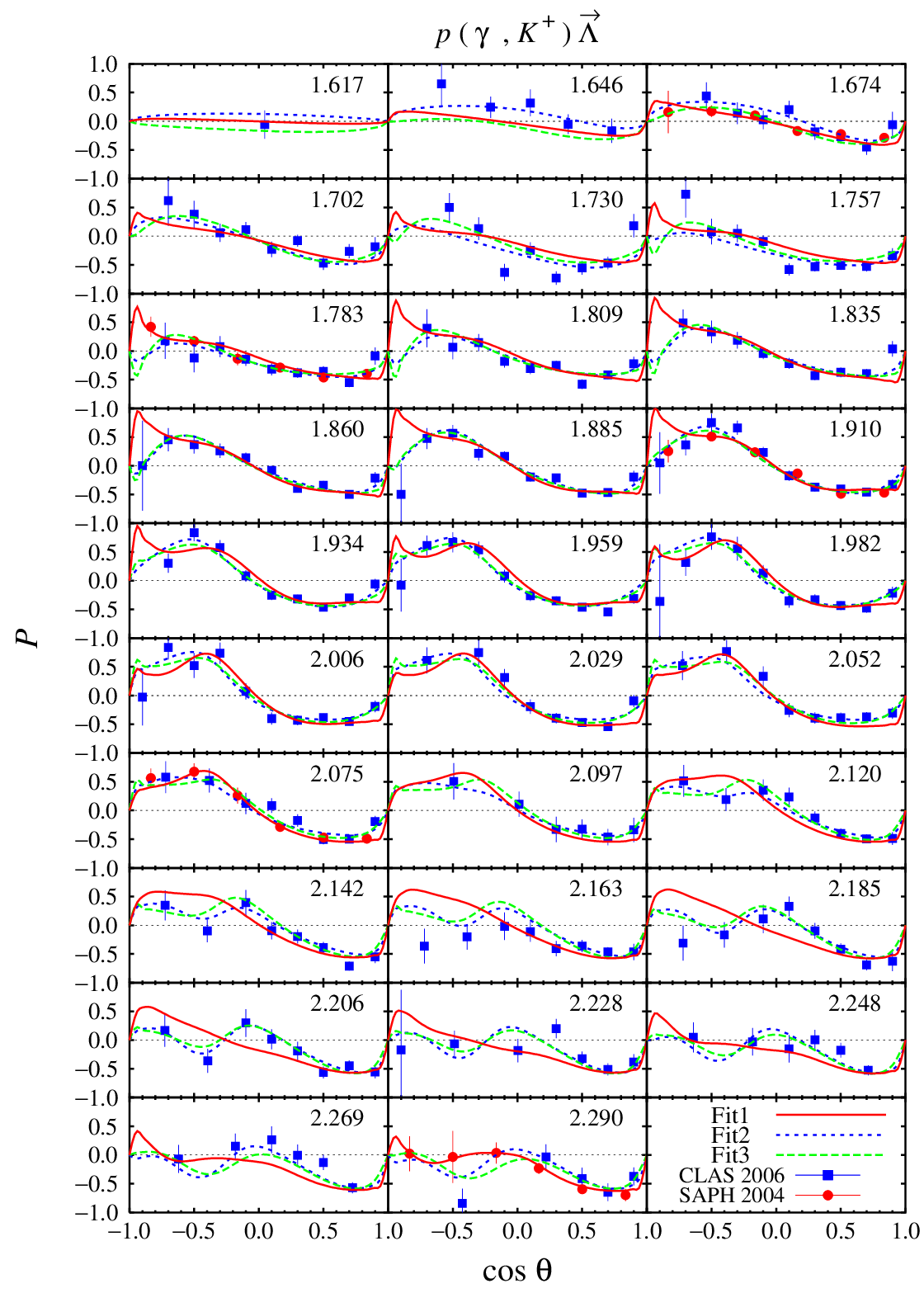,width=12cm}}
\caption{(Color online) Comparison between the $\Lambda$ recoil polarization  
  obtained from the three fits with CLAS and SAPHIR data.}\label{fig:clas_pollam}
\end{figure}   

\begin{figure*}[!]
\centering
 \mbox{\epsfig{file=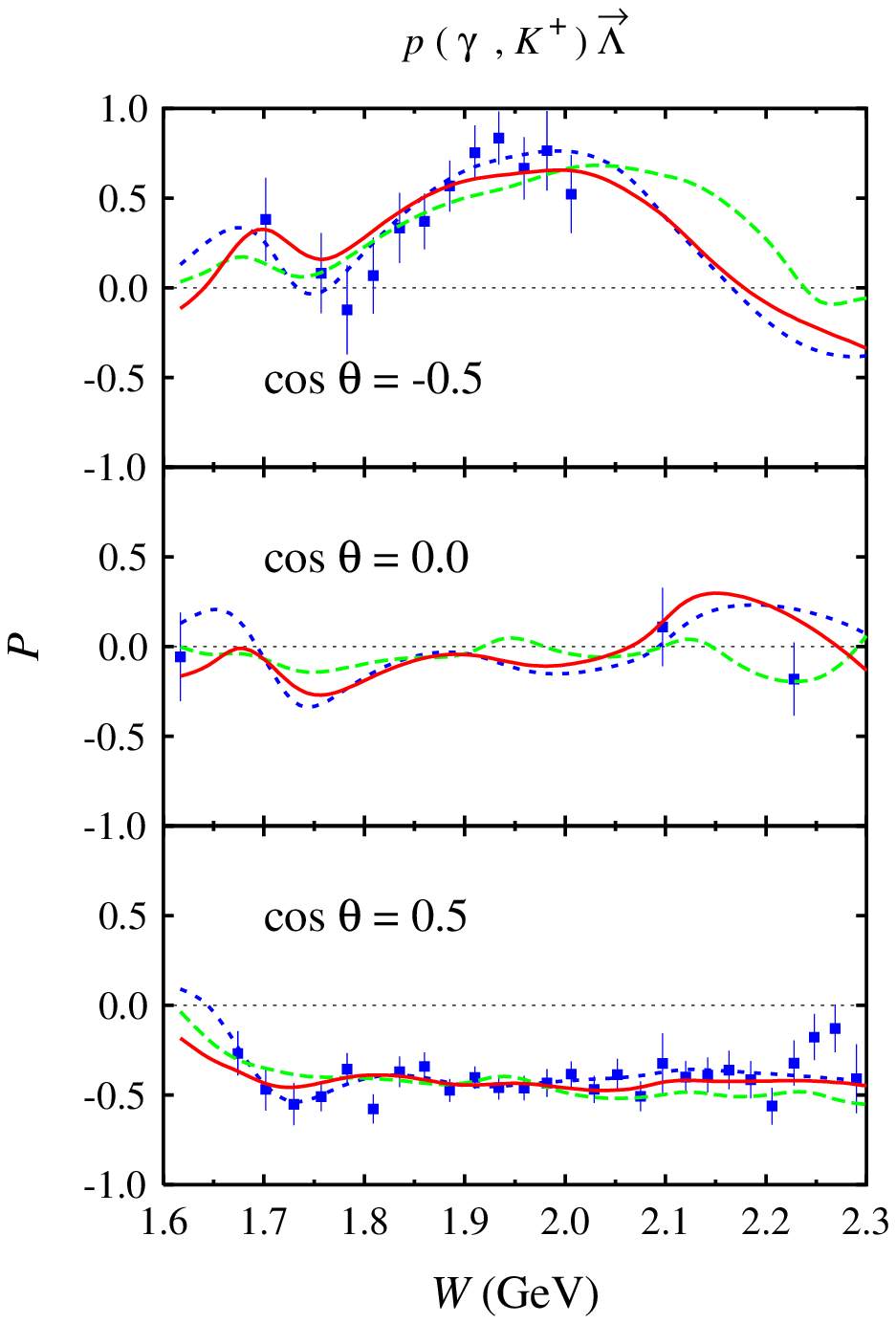,width=7cm}}
\caption{(Color online) Same as in Fig.~\ref{fig:clas_pollam}, but
  for the $W$ distributions.}\label{fig:clas_pollam_w}
\end{figure*}

The $\Lambda$ recoil polarizations obtained from all fits are compared 
with experimental data in Fig.~\ref{fig:clas_pollam}. Except at higher 
energies and in backward directions, where experimental data  have
large error bars, no result shows any significant difference.
Therefore, in view of the present error bars, the $\Lambda$ recoil polarization
is not a decisive observable for revealing further information from the three fits. 
The energy distribution of this observable shown in Fig.~\ref{fig:clas_pollam_w} 
emphasizes this argument. At $\cos\theta = 0.5$ it is interesting to remark that
the polarizations predicted by the three fits are almost similar and
the values are almost constant at about $-0.5$ over the whole energy range, 
except very close to threshold.

Both CLAS and SAPHIR collaborations extracted the total cross sections 
and displayed them graphically. The numerical data points shown
in Fig.~\ref{fig:total} were taken from total cross section figures of
Refs.\,\cite{Glander:2003jw,Bradford:2005pt} and, for the sake of 
consistency,  not used in the fits.
In Ref.~\cite{Julia-Diaz:2006is} it was suspected that the
two collaborations have extracted the total cross sections in different
ways, hence the discrepancy between them seems to be larger than that in
differential cross sections. However, by comparing the solid line
and solid squares, as well as the dotted line and solid circles in
Fig.~\ref{fig:total}, we conclude that the extracted total cross
sections from both collaborations are consistent with their differential
cross sections. The fact that the discrepancy is more profound in the
total cross sections is seemingly due to the cumulative effect of the integration,
which can be immediately comprehended if we compare the solid lines (fit to
the SAPHIR data) with dotted lines (fit to the CLAS data) in Fig.~\ref{fig:clas_dkpl_e}.
On the other hand, the result of Fit 3 (dashed line in Fig.~\ref{fig:total})
clearly indicates that including both data sets in the fit results in a model
which is consistent with no data set, as has been previously pointed out  
by Ref.~\cite{Bydzovsky:2006wy}. 

\begin{figure*}[!]
\centering
 \mbox{\epsfig{file=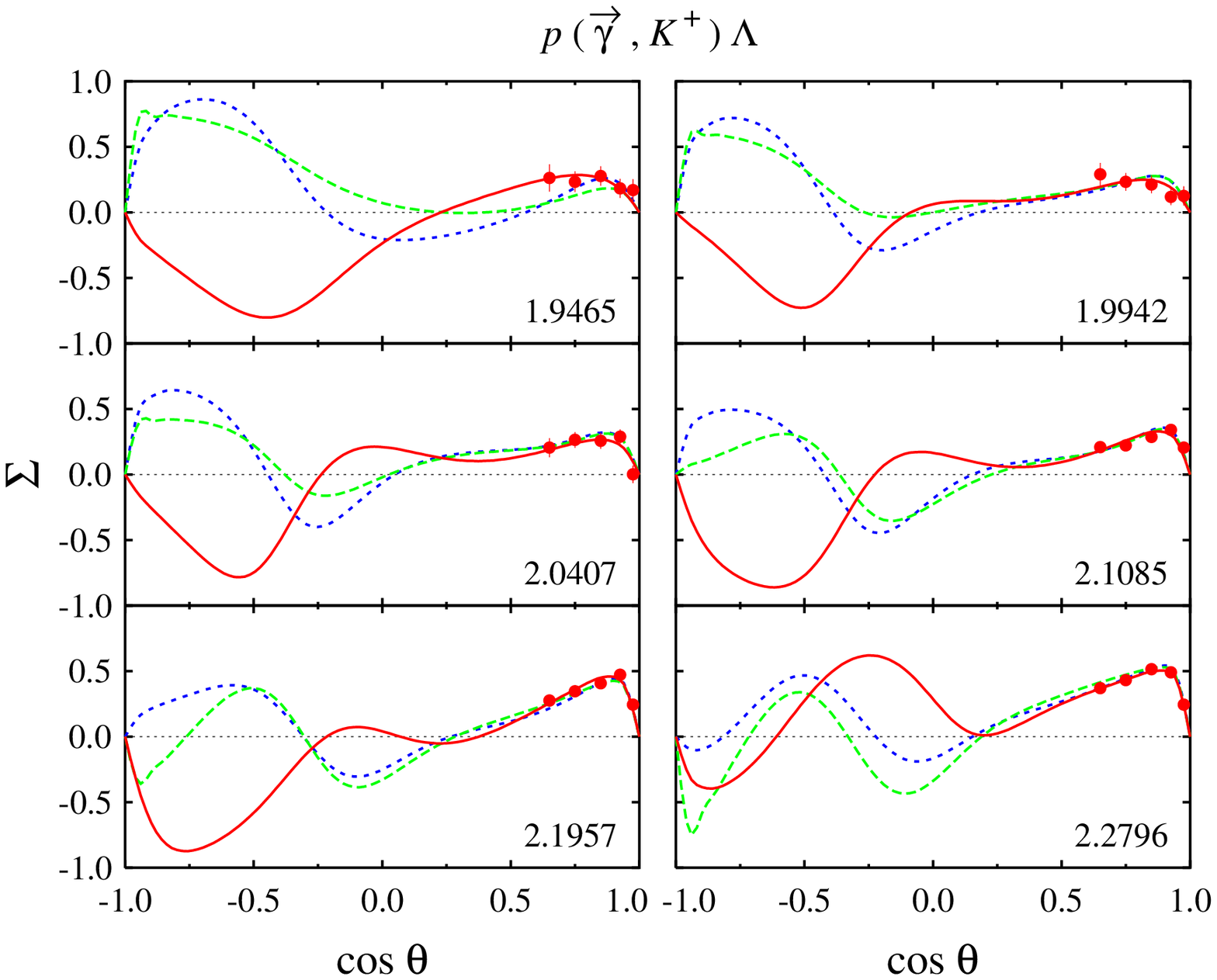,width=12cm}}
\caption{(Color online) Photon asymmetry obtained from the three fits compared 
  with the LEPS data. Notation for the curves is the same as in 
  Fig.~\ref{fig:clas_dkpl}.}\label{fig:polpho}
\end{figure*}

The result for the polarized photon beam asymmetry is shown in Fig.~\ref{fig:polpho},
where we can obviously see a good agreement between predictions of all fits
and the experimental data from LEPS. 
This result also corroborates the finding of Ref.~\cite{Julia-Diaz:2006is}
that further measurements of this observable in the backward directions
would put a strong constraint on the model. Reference\,\cite{Julia-Diaz:2006is} 
found that the currently available experimental data of this observable
(see the last line of Table~\ref{tab:chi}) generate about 13\% of the 
total $\chi^2$. In contrast to this, we found that the data contribute 
only 3\% (2\%) to the total $\chi^2$ of Fit 1 (Fit 2), which shows 
a better agreement of the three fits compared to models M1 and M2
of Ref.~\cite{Julia-Diaz:2006is}.

\begin{figure*}[!]
\centering
 \mbox{\epsfig{file=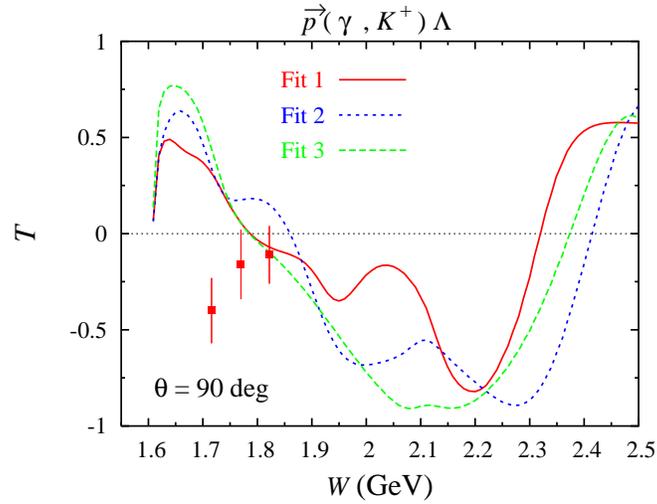,width=9cm}}
\caption{(Color online) Comparison between target asymmetries obtained 
  from the three different fits and experimental data \cite{Althoff:1978qw}. 
  Note that the three data points were
  not used in the fits.}\label{fig:target}
\end{figure*}

In Fig.~\ref{fig:target} we compare the target asymmetry predicted by 
the three fits with experimental data. There are only three data points
with large error bars available for this observable. To our knowledge,
except for the Saclay-Lyon model \cite{David:1995pi} which is only in fair agreement
with those data, other models \cite{Julia-Diaz:2006is,Maxwell:2004ga,Mart:1999ed}
fail to reproduce them. Figure~\ref{fig:target} clearly indicates that
this observable could provide a stringent constraint to phenomenological
models that try to explain the $K^+\Lambda$ photoproduction process. At this
stage, it is important to note that we also obtained another solution for
Fit 2 which can nicely reproduce these three data points. However, since
the extracted parameters are quite different from those of Fit 1 and Fit 3
(as well as from the PDG values) and the corresponding $\chi^2=1610$ is slightly
larger than that of Fit 2 (i.e., $\chi^2=1581$), we do not follow up 
 this alternative. Moreover, the number of available data and the
size of error bars make it difficult to draw a firm conclusion on the 
discrepancy shown in Fig.~\ref{fig:target}. Future measurement
of this asymmetry from threshold up to $W\approx 2.2$ GeV with error bars
comparable to those of CLAS and SAPHIR data would certainly help to
clarify this issue.

\begin{figure*}[hbt]
\centering
 \mbox{\epsfig{file=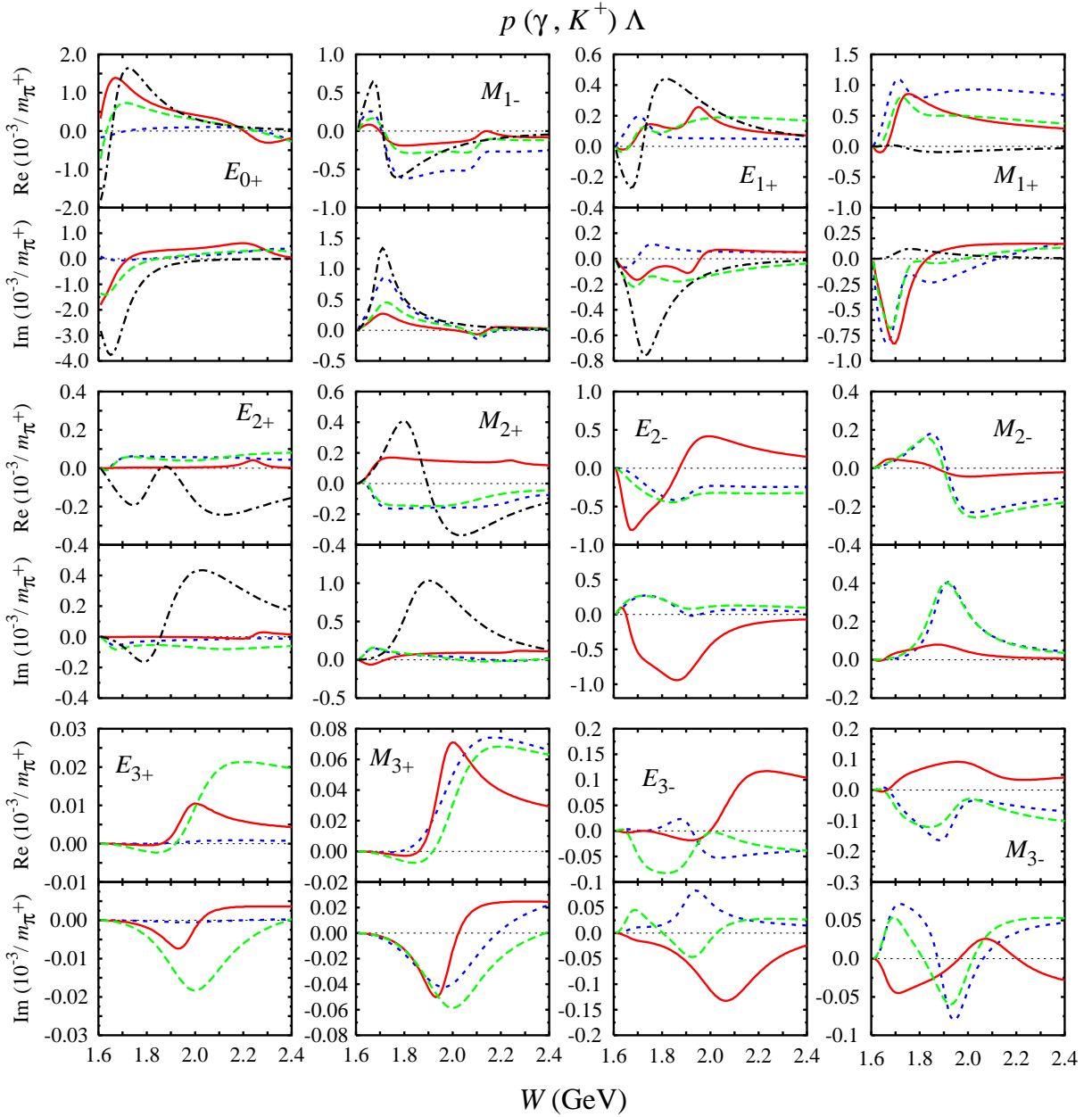,width=16cm}}
\caption{(Color online) Electric and magnetic multipoles obtained from
  the three fits. The dash-dotted curves show the prediction of 
  Kaon-Maid. Notation for other curves is the same as in Fig.~\ref{fig:clas_dkpl}. }
\label{fig:multipole}
\end{figure*}

In Fig.~\ref{fig:multipole} we display the multipole amplitudes for $\ell \le 3$.
It is obvious from this figure that the predicted multipoles are different in most
cases, as has been also pointed out by Ref.~\cite{Mart:2004ug}. This again 
indicates that the problem of mutual consistency in the presently available
data still prohibits a more model-independent multipole analysis in kaon
photoproduction.

\subsection{The First Peak at $W\approx 1.7$ GeV}

\begin{figure}[!]
\centering
 \mbox{\epsfig{file=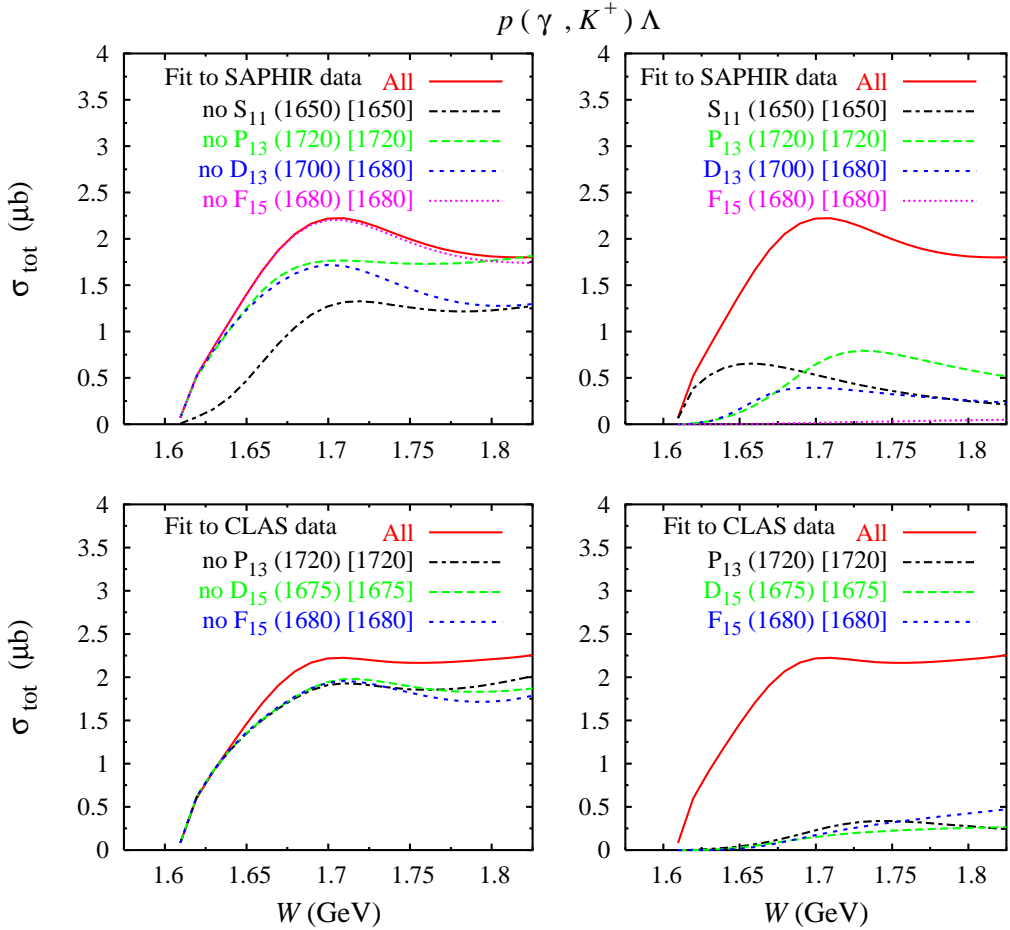,width=14cm}}
\caption{(Color online) Contribution from resonances with masses around 1700 MeV
  to the total cross sections in the case of Fit 1 (upper panels) and Fit 2 (lower 
  panels). For comparison, values of the extracted masses are shown in the square 
  brackets.}
\label{fig:first-peak}
\end{figure}   

Both SAPHIR and CLAS data show an obvious peak at $W$ around 1.7 GeV
in the total as well as differential cross sections.
Since these two data sets seem to be similar up to $W\approx 1.7$ GeV,
it is reasonable to expect a similar resonance behavior below this energy region.
However, Table~\ref{tab:results_resonance} indicates that, except for the
$P_{13}(1720)$ and $F_{15}(1680)$, the two data sets require different
resonances. This is elucidated in Fig.~\ref{fig:first-peak}, where we compare
the contribution of relevant resonances with masses around 1.7 GeV 
to the total cross section.

Due to their large $\Delta\chi^2$, contributions from the $S_{11}(1650)$, 
$P_{13}(1720)$, and $D_{13}(1700)$ in the case of Fit 1 
(the two upper panels of Fig.~\ref{fig:first-peak}) 
are easily comprehended. The $F_{15}(1680)$ contribution, which is according
to Table~\ref{tab:results_resonance} is also important, is found to be 
important to describe the SAPHIR data only at the very forward angles. As a 
consequence, its contribution is difficult to see in this figure.

In contrast to the previous case, contributions of these resonances are
somewhat complicated in the case of Fit 2  (the two lower panels of 
Fig.~\ref{fig:first-peak}). This is mainly due to the relatively large 
background of the Fit 2 (see Fig.~\ref{fig:born}). Nevertheless, contributions
from the $P_{13}(1720)$, $D_{15}(1675)$, and  $F_{15}(1680)$ are still
sizable. These contributions are required to decrease the cross 
section down to the experimental value through destructive interference.

The above result is clearly unexpected. However, we can understand this
by carefully examine the total cross section data shown in 
Fig.~\ref{fig:total} or the differential cross section data shown in
Figs.~\ref{fig:clas_dkpl_e} and \ref{fig:saph_dkpl_e}, where we can see that at 
$W=1.7$ GeV the discrepancy between the two data sets starts to appear.
Given that the lowest lying resonance used in this analysis is the $S_{11}(1650)$,
which has a width of 150 MeV, all experimental data up to $W=1.8$ GeV 
will certainly influence the extracted resonance parameters.

Another possible origin of the above finding is that the two data sets 
are already different for $W\lesssim 1.7$ GeV. To investigate this, we 
separately fitted both SAPHIR and CLAS differential cross sections data from
threshold up to $W\approx 
1.7$ GeV, by including the $S_{11}(1650)$, $P_{11}(1710)$, $P_{13}(1720)$, 
$D_{13}(1700)$, $D_{15}(1675)$, and $F_{15}(1680)$ resonances. We found
that the extracted resonance parameters from the two fits are quite different, 
which, therefore, confirms that the two data sets are already different 
at $W\lesssim 1.7$ GeV.

\subsection{The Second Peak at $W\approx 1.9$ GeV}
For almost one decade since the previous SAPHIR data were published in 1998 
\cite{Tran:1998qw} there has been a lot of discussion on which resonance
is responsible for explaining the second peak at $W\approx 1.9$ GeV in
the total as well as differential
cross sections. Here, it is important to note that, although varying as a function
of the kaon angle in the latter case, the peak still exists in both CLAS
and SAPHIR data. 

The debate was ignited by the authors of Ref.~\cite{Mart:1999ed},
who, by means of the results from a certain constituent quark model
\cite{capstick94} and an isobar model, interpreted the peak as the 
existence of the missing resonance $D_{13}(1895)$. Subsequently, 
it was shown by Janssen {\it et al.} \cite{Janssen:2001wk} that
the peak could be also equally well reproduced by including a $P_{13}(1950)$
resonance. However, most of analyses based on the isobar model after that
confirmed that including the $D_{13}(1895)$ will significantly improve the
agreement with experimental data \cite{d13_support}. 

A recent partial wave analysis by Anisovich {\it et al.} 
\cite{Anisovich:2005tf} found that a new $D_{13}$ with $M=1875\pm 25$ 
MeV and $\Gamma = 80\pm 20$ MeV is needed in order
to explain the processes $\gamma p\to \pi N, \eta N, K\Lambda$ and $K\Sigma$.
Experimental data on the $\gamma p\to N^*(\Delta^*) \to \pi^0 p$ published by
CB-ELSA collaboration not long after that shifted this resonance
to a higher mass, i.e., $M=1943\pm 17$ MeV and $\Gamma = 82\pm 20$ MeV
\cite{Bartholomy:2004uz}. 
By analyzing the new SAPHIR data within a multipole approach Ref.~\cite{Mart:2004ug}
found that the $D_{13}$ could have a mass and width of $1912$ MeV and 148 MeV
(Model II of Table 1 in Ref.~\cite{Mart:2004ug}). Meanwhile, 
a very recent coupled-channel analysis for the $\pi N\to KY$ and 
$\gamma N\to KY$ processes puts this resonance at $M=1912$ MeV (or $1954$ MeV) and
$\Gamma = 316$ MeV (or $249$ MeV), depending on the data set used in the fit
\cite{Julia-Diaz:2006is}. 
Therefore, the obvious question is
whether or not the second peak near $W\approx 1900$ MeV signals a $D_{13}$ resonance
with a mass of around 1900 MeV. 

\begin{figure*}[!]
\centering
 \mbox{\epsfig{file=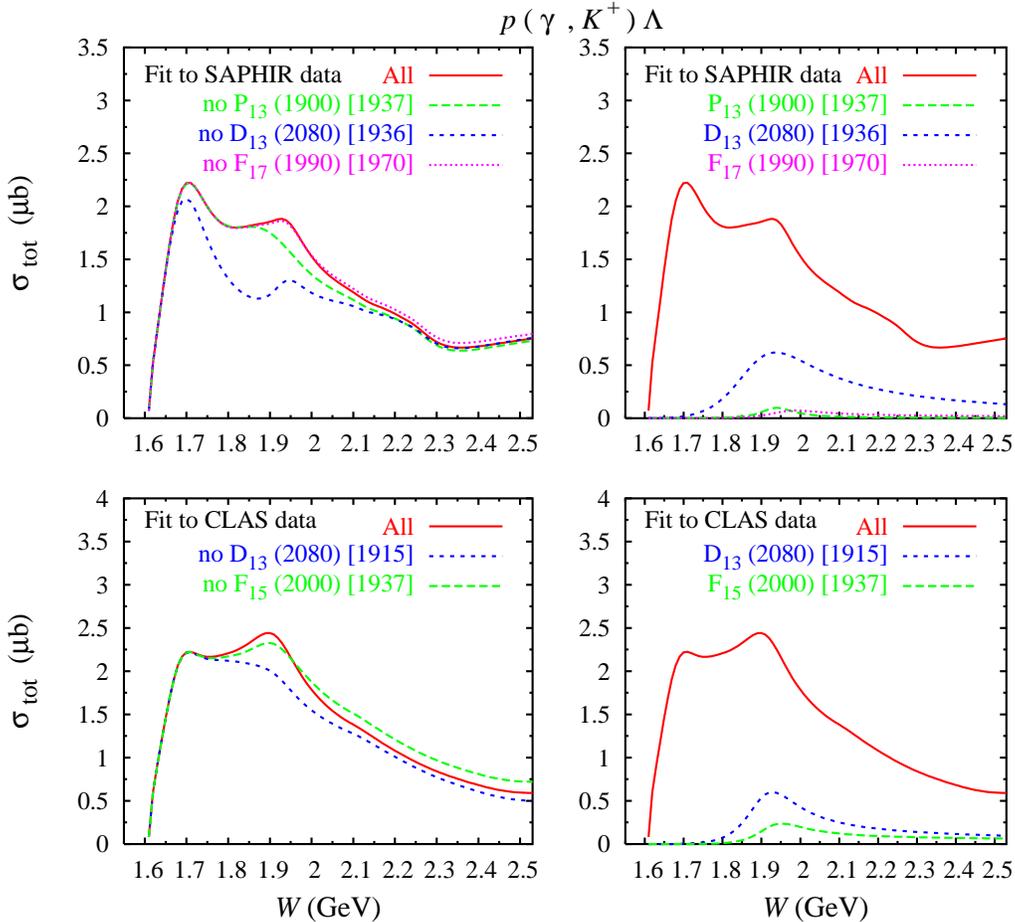,width=14cm}}
\caption{(Color online) Same as Fig.~\ref{fig:first-peak}, except 
  for the resonances with masses around 1900 MeV, which contribute to 
  the second peak in the total cross sections.}
\label{fig:contribute}
\end{figure*}   

To answer this question let us look at Fig.~\ref{fig:contribute}, 
where we show the comparison between total cross sections of both Fit 1 and Fit 2 
obtained by including all resonances and those obtained by 
excluding resonances with masses around 1900 MeV in 
the left panels. In the right panels a comparison between total cross sections 
obtained by including all resonances and those obtained from the individual 
resonances is shown. In the case of
Fit 1 (upper panels), it is obvious that the $D_{13}(2080)$ with a mass of 
1936 MeV provides the dominant contribution to this second peak. This can
also immediately be seen from Fig.~\ref{fig:strength} or from 
Table \ref{tab:results_resonance}, where we see that the corresponding 
$\Delta\chi^2=8.8\%$ is larger than that of the $P_{13}(1900)$ (4.4\%), 
or the $F_{17}(1990)$ (2.7\%). Albeit using a different formalism, 
this result is consistent with our previous finding \cite{Mart:2004ug}, as well as 
with various analyses \cite{Bartholomy:2004uz,Julia-Diaz:2006is}.
The reason that the mass of this $D_{13}$ is shifted toward a higher
value compared with the previous observation (1895 MeV as obtained
in Ref.~\cite{Mart:1999ed}) seemingly originates from the new
SAPHIR data \cite{Glander:2003jw} which have the second peak at higher 
$W$ compared with the previous ones \cite{Tran:1998qw}
(see Fig.~\ref{fig:saph_dkpl_e}).

Interestingly, as shown by Fig.~\ref{fig:strength} and 
Table~\ref{tab:results_resonance}, the new CLAS data yield the same 
conclusion. Using this data set (Fit 2) the extracted mass 
of $D_{13}$ is 1915 MeV, which is very close to the value given by Fit 1
(1936 MeV). As shown by Fig.~\ref{fig:strength} this
resonance appears to be quite decisive in the process
($\Delta\chi^2=8.5$\%), and from the lower-left panel of 
Fig.~\ref{fig:contribute} it is obvious that excluding this resonance 
in the process drastically changes the shape of the cross section. 
We also note that including all data sets in the fit does not
change this conclusion.

To summarize this subsection we may say that within this 
multipole approach the two data sets lead
to the same conclusion on the origin of the second peak in the
$W$ distribution of the cross sections, i.e., the $D_{13}(2080)$ with
a mass between 1911 -- 1936 MeV.

\section{Conclusions and Outlook}\label{sec:conclusion}
We have analyzed the $\gamma p \to K^+\Lambda$ process by means of a multipole
approach with a gauge-invariant, crossing-symmetric background amplitude
obtained from tree-level Feynman diagrams. The corresponding free parameters
are fitted to three different data sets, i.e., combinations of SAPHIR and 
LEPS data, CLAS and LEPS data, and all of these data. Results of the fit
indicate the lack of mutual consistency between SAPHIR and CLAS data,
whereas the LEPS data are shown to be more consistent with the CLAS ones.
In most cases, the extracted parameters from the three data sets 
are found to be different and, therefore, could lead
to different conclusions if those data were used individually or simultaneously
to extract the information on missing resonances. 

From a fit to SAPHIR and LEPS data it is found that the $S_{11}(1650)$, 
$P_{13}(1720)$, $D_{13}(1700)$, $D_{13}(2080)$, $F_{15}(1680)$, and $F_{15}(2000)$
resonances are more important than other resonances used in this analysis, 
whereas fitting to the combination of CLAS and LEPS data indicates that the 
$P_{13}(1900)$, $D_{13}(2080)$, $D_{15}(1675)$, $F_{15}(1680)$, and $F_{17}(1990)$ 
resonances 
to be more decisive ones. It is shown that fitting to all data simultaneously
changes this conclusion and results in a model which is inconsistent to
all data sets.

Our analysis indicates that the target asymmetry cannot be described by any 
of the models. In view of the current available experimental data we conclude
that measurement of this observable should be addressed in a 
future experimental proposal.

The three-star resonance $P_{11}(1710)$ that has been used in almost all 
isobar models within both single-channel and multi-channel approaches is
found to be insignificant to the $K^+\Lambda$ photoproduction
by both SAPHIR and CLAS data.

It is also found that the second peak in cross sections at $W\sim 1900$ MeV
is originated from the $D_{13}(2080)$  resonance. The extracted mass 
would be 1936 MeV if SAPHIR data were used
or 1915 MeV if CLAS data were used. This finding would not change if
all data sets were used.

We have observed that the total cross sections reported by the two collaborations
are consistent with their differential cross sections. The fact that the
discrepancy is larger in the total cross sections stems from
the cumulative effect of the integration. 

Although results of the present work could reveal certain consequences
of using SAPHIR or CLAS data in the database, it is still difficult to
determine which data set should be used in order to obtain the correct
resonance parameters. We also realize that the results presented here
are not final, because a more representative calculation should ideally 
be performed in a 
coupled-channels formalism where other channels such as $\pi N$, $\eta N$,
$\pi\pi N$, and $\omega N$ are also taken into account. Nevertheless,
the simple calculation presented here has revealed two most important
issues that will need to be addressed in future calculations: 
(1) contribution from higher spin resonances are important,
(2) until we can settle the problem of data consistency, the results 
of all calculations are now data dependent. Future measurements 
such as the one planned at MAMI in Mainz are, therefore, expected 
to remedy this unfortunate situation. 

Our next goal is to consider the $\gamma p\to K^+\Sigma^0$ channel and
to incorporate the effect of other channels.

\section*{Acknowledgment}
The authors thank William J. Briscoe for carefully reading the manuscript
and acknowledge the support from the Faculty of
Mathematics and Sciences, University of Indonesia, as well as 
from the Hibah Pascasarjana grant.

\renewcommand{\baselinestretch}{1.5}


\begin{thebibliography}{0}
\bibitem{Adelseck:1986fb}
  R.~A.~Adelseck, C.~Bennhold and L.~E.~Wright, Phys.\ Rev.\ C {\bf 32}, 1681 (1985).
\bibitem{Adelseck:1990ch}
  R.~A.~Adelseck and B.~Saghai, Phys.\ Rev.\ C {\bf 42}, 108 (1990).
\bibitem{Williams:1991tw}
  R.~A.~Williams, C.~R.~Ji and S.~R.~Cotanch, Phys.\ Rev.\ C {\bf 43}, 452 (1991).
\bibitem{Han:1999ck}
  B.~S.~Han, M.~K.~Cheoun, K.~S.~Kim and I.~T.~Cheon, Nucl.\ Phys.\ A {\bf 691}, 713 (2001).
\bibitem{kaon-maid} T. Mart, C. Bennhold, H. Haberzettl, and L. Tiator, Kaon-Maid, 
  available at http://www.kph.uni-mainz.de/MAID/kaon/kaonmaid.html. 
  The published versions are available in: \cite{Mart:1999ed};
  T.~Mart, Phys.\ Rev.\ C {\bf 62}, 038201 (2000); C.~Bennhold,
  H.~Haberzettl and T.~Mart, arXiv:nucl-th/9909022.
\bibitem{Feuster:1998cj}
  T.~Feuster and U.~Mosel, Phys.\ Rev.\ C {\bf 59}, 460 (1999).
\bibitem{Julia-Diaz:2006is}
  B.~Julia-Diaz, B.~Saghai, T.~S.~Lee and F.~Tabakin, Phys.\ Rev.\ C 
  {\bf 73}, 055204 (2006).
\bibitem{Chiang:2001pw}
  W.~T.~Chiang, F.~Tabakin, T.~S.~H.~Lee and B.~Saghai,
  Phys.\ Lett.\ B {\bf 517}, 101 (2001).
\bibitem{Li:1995sia}
  Z.~P.~Li, Phys.\ Rev.\ C {\bf 52}, 1648 (1995).
\bibitem{Lu:1995bk}
  D.~H.~Lu, R.~H.~Landau and S.~C.~Phatak, Phys.\ Rev.\ C {\bf 52}, 1662 (1995).
\bibitem{Mart:2003yb}
  T.~Mart and T.~Wijaya, Acta Phys.\ Polon.\ B {\bf 34}, 2651 (2003).
\bibitem{Mart:2004au}
  T.~Mart and C.~Bennhold,``Kaon photoproduction in the Feynman and Regge theories,''
  arXiv:nucl-th/0412097.
\bibitem{Corthals:2005ce}
  T.~Corthals, J.~Ryckebusch and T.~Van Cauteren, Phys.\ Rev.\ C {\bf 73}, 045207 (2006).
\bibitem{David:1995pi}
  J.~C.~David, C.~Fayard, G.~H.~Lamot and B.~Saghai, Phys.\ Rev.\ C {\bf 53}, 2613 (1996).
\bibitem{Renard:1971us}
  F.~M.~Renard and Y.~Renard, Nucl.\ Phys.\ B {\bf 25}, 490 (1971);
  Y.~Renard, Nucl.\ Phys.\ B {\bf 40}, 499 (1972);
  Y. Renard, Th\`ese de Doctorat d'Etat \`es-Sciences 
  Physiques, Universit\'e des Sciences et Techniques du 
  Languedoc, 1971 (unpublished).
\bibitem{Eidelman:2004wy} S.~Eidelman {\it et al.}  [Particle Data Group],
  Phys.\ Lett.\ B {\bf 592}, 1 (2004).
\bibitem{Glander:2003jw}
  K.~H.~Glander {\it et al.},
  Eur.\ Phys.\ J.\ A {\bf 19}, 251 (2004).
\bibitem{Bradford:2005pt}
  R.~Bradford {\it et al.}  [CLAS Collaboration],
  Phys.\ Rev.\ C {\bf 73}, 035202 (2006).
\bibitem{Bydzovsky:2006wy}
  P.~Bydzovsky and T.~Mart, 
  ``Analysis of the data consistency on kaon photoproduction with Lambda in the
  final state,''
  arXiv:nucl-th/0605014.
\bibitem{Mart:2004ug}
  T.~Mart, A.~Sulaksono and C.~Bennhold, arXiv:nucl-th/0411035; also in \cite{d13_support}.
\bibitem{hanstein99} D.~Drechsel, O.~Hanstein, S.~S.~Kamalov and L.~Tiator,
  Nucl.\ Phys.\ A {\bf 645}, 145 (1999); [arXiv:nucl-th/9807001].
\bibitem{Tiator:2003uu}
  L.~Tiator, D.~Drechsel, S.~Kamalov, M.~M.~Giannini, E.~Santopinto and A.~Vassallo,
  Eur.\ Phys.\ J.\ A {\bf 19}, 55 (2004);
  [arXiv:nucl-th/0310041].
\bibitem{background} The explicit expression for the background amplitudes 
  are given, e.g., in Ref.~\cite{Mart:2003yb}, or in 
  T.~Mart, Ph.D Thesis, Universit\"at Mainz, 1996 (unpublished).
\bibitem{thom1966} H. Thom, Phys. Rev. {\bf 151}, 1322 (1966).
\bibitem{Adelseck:yv} R.~A.~Adelseck and L.~E.~Wright, 
  Phys.\ Rev.\ C {\bf 38}, 1965 (1988).
\bibitem{ohta89}
  K. Ohta, Phys. Rev.\ C {\bf 40}, 1335 (1989).
\bibitem{Davidson:2001rk} R.~M.~Davidson and R.~Workman, Phys.\ Rev.\ C 
  {\bf 63}, 025210 (2001).
\bibitem{Haberzettl:1998eq}
H.~Haberzettl, C.~Bennhold, T.~Mart and T.~Feuster, Phys.\ Rev.\ C {\bf 58}, 40 (1998).
\bibitem{Chiang:2001as}
  W.~T.~Chiang, S.~N.~Yang, L.~Tiator and D.~Drechsel,
  Nucl.\ Phys.\ A {\bf 700}, 429 (2002).
\bibitem{Aznauryan:2002gd}
  I.~G.~Aznauryan,
  Phys.\ Rev.\ C {\bf 67}, 015209 (2003).
\bibitem{Knochlein:1995qz} G.~Knochlein, D.~Drechsel 
  and L.~Tiator, Z.\ Phys.\ A {\bf 352}, 327 (1995).
\bibitem{McNabb2004} J.~W.~C.~McNabb {\it et al.}  [The CLAS Collaboration],
  Phys.\ Rev.\ C {\bf 69}, 042201 (2004); J.~W.~C.~McNabb, PhD Thesis, Carnegie Mellon
  University (2002); R. Schumacher, private communication.
\bibitem{Sumihama:2005er}
  M.~Sumihama {\it et al.}  [LEPS Collaboration],
  Phys.\ Rev.\ C {\bf 73}, 035214 (2006).
\bibitem{Althoff:1978qw}
  K.~H.~Althoff {\it et al.},
  Nucl.\ Phys.\ B {\bf 137}, 269 (1978).
\bibitem{Maxwell:2004ga}
  O.~V.~Maxwell, Phys.\ Rev.\ C {\bf 70}, 044612 (2004).
\bibitem{Mart:1999ed}
  T.~Mart and C.~Bennhold, Phys.\ Rev.\ C {\bf 61}, 012201 (2000).
\bibitem{older-measurment} References for old measurements are listed in 
  Table IX of \cite{Adelseck:1990ch}, or in references of \cite{Tran:1998qw}.
\bibitem{mart-tiator} T. Mart and L. Tiator, work in progress.
\bibitem{Tran:1998qw}
  M.~Q.~Tran {\it et al.}  [SAPHIR Collaboration], Phys.\ Lett.\ B {\bf 445}, 20 (1998).
\bibitem{capstick94} S. Capstick and W. Roberts, Phys. Rev. D {\bf 49},
  4570 (1994); Phys. Rev. D {\bf 58}, 074011
  (1998); S. Capstick, Phys. Rev. D {\bf 46}, 2864 (1992).
\bibitem{Janssen:2001wk}
  S.~Janssen, J.~Ryckebusch, D.~Debruyne and T.~Van Cauteren,
  Phys.\ Rev.\ C {\bf 65}, 015201 (2001).
\bibitem{d13_support} See e.g.: T.~K.~Choi, M.~K.~Cheoun, K.~S.~Kim and B.~G.~Yu, in 
  Proceedings of the International Symposium On Electrophoto-Production 
  Of Strangeness On Nucleons And Nuclei (SENDAI 03), 16-18 Jun 2003, Sendai, Japan, 
  edited by K. Maeda, H. Tamura, S.N. Nakamura, O. Hashimoto. River Edge, 
  World Scientific, 2004, pp. 85.
\bibitem{Anisovich:2005tf}
  A.~V.~Anisovich, A.~Sarantsev, O.~Bartholomy, E.~Klempt, V.~A.~Nikonov and U.~Thoma,
  Eur.\ Phys.\ J.\ A {\bf 25}, 427 (2005).
\bibitem{Bartholomy:2004uz}
  O.~Bartholomy {\it et al.} [CB-ELSA Collaboration],
  Phys.\ Rev.\ Lett.\  {\bf 94}, 012003 (2005).
\end{thebibliography}
\end{document}